\documentclass[%
 reprint,
nofootinbib,
 amsmath,amssymb,
 aps, prl
]{revtex4-1}

\usepackage[dvips]{graphicx}
\usepackage{latexsym,epsfig,bm,psfrag}
\usepackage{subfigure}
\usepackage{color}         
\usepackage{dcolumn}                 
\usepackage[bookmarksnumbered,bookmarksopen,colorlinks,citecolor=blue,linkcolor=blue]{hyperref} %
\usepackage{calrsfs}
\DeclareMathOperator{\sgn}{sgn}

\begin{document}
\renewcommand{\vec}{\boldsymbol}
\newcommand{\mc}{M_{\mathrm{c}}}
\newcommand{\mn}{M_{\mathrm{n}}}
\newcommand{\mnc}{M_{\mathrm{nc}}} 
\newcommand{\mr}{M_{_{\mathrm{R}}}}
\newcommand{\gma}{\gamma}
\newcommand{\gmat}{\tilde{\gamma}}
\newcommand{\pc}{\vec{p}_{c}}
\newcommand{\pn}{\vec{p}_{n}}
\newcommand{\bes}{{}^{7}\mathrm{Be}}
\newcommand{\be}{{}^{8}\mathrm{B}}
\renewcommand{\S}[2]{{}^{#1}S_{#2}}
\renewcommand{\P}[2]{{}^{#1}P_{#2}}
\newcommand{\gone}{g_{(\S{3}{1})}}
\newcommand{\gtwo}{g_{(\S{5}{2})}}
\newcommand{\gthree}{g_{(\S{3}{1}^{*})}}
\newcommand{\aone}{a_{(\S{3}{1})}}
\newcommand{\atwo}{a_{(\S{5}{2})}}
\newcommand{\hone}{h_{(\P{3}{2})}}
\newcommand{\htwo}{h_{(\P{5}{2})}}
\newcommand{\hpt}{h_{Pt}}
\newcommand{\hthree}{h_{(\P{3}{2}^{*})}}
\newcommand{\honet}{\tilde{h}_{(\P{3}{1})}}
\newcommand{\htwot}{\tilde{h}_{(\P{5}{1})}}
\newcommand{\Xone}{{X}_{(\S{3}{1})}}
\newcommand{\Xtwo}{{X}_{(\S{5}{2})}}
\newcommand{\Xonet}{\tilde{X}_{(\S{3}{1})}}
\newcommand{\Xtwot}{\tilde{X}_{(\S{5}{2})}}
\newcommand{\V}[1]{\vec{V}_{#1}}
\newcommand{\fdu}[2]{{#1}^{\dagger #2}}
\newcommand{\fdd}[2]{{#1}^{\dagger}_{#2}}
\newcommand{\fu}[2]{{#1}^{#2}}
\newcommand{\fd}[2]{{#1}_{#2}}
\newcommand{\T}[2]{T_{#1}^{\, #2}}
\newcommand{\e}{\vec{\epsilon}}
\newcommand{\es}{\e^{*}}
\newcommand{\cw}[2]{\chi^{(#2)}_{#1}}
\newcommand{\cwc}[2]{\chi^{(#2)*}_{#1}}
\newcommand{\cwf}[1]{F_{#1}}
\newcommand{\cwg}[1]{G_{#1}}
\newcommand{\ke}{k_{E}}
\newcommand{\kest}{k_{E\ast}}
\newcommand{\kc}{k_{C}}
\newcommand{\upartial}[1]{\partial^{#1}}
\newcommand{\dpartial}[1]{\partial_{#1}}
\newcommand{\etae}{\eta_{E}}
\newcommand{\etab}{\eta_{B}}
\newcommand{\etaest}{\eta_{E\ast}}
\newcommand{\etabst}{\eta_{B\ast}}
\newcommand{\vecpt}[1]{\hat{\vec{#1}}}
\newcommand{\uY}[2]{Y_{#1}^{#2}}
\newcommand{\dY}[2]{Y_{#1 #2}}
\newcommand{\abi}{{\it ab initio} \ }
\newcommand{\half}{\frac{1}{2}}
\newcommand{\threehalf}{\frac{3}{2}}
\newcommand{\Gaf}[1]{\Gamma\!\left(#1\right)}
\newcommand{\eq}[1]{Eq.~(\ref{#1})}
\newcommand{\eqp}[1]{Equation~(\ref{#1})}
\newcommand{\eqtoeq}[2]{Eqs.~(\ref{#1})--(\ref{#2})}
\newcommand{\eqtoeqp}[2]{Equations~(\ref{#1})--(\ref{#2})}
\newcommand{\ze}{z_{_{E}}}
\newcommand{\pe}{p}
\newcommand{\eL}{E_{_L}}
\newcommand{\mypsi}{\mathcal{B}}
\newcommand{\MH}{M_H}
\newcommand{\ith}{i^{\textrm{th}}}
\newcommand{\jth}{j^{\textrm{th}}}
\newcommand{\lth}{\ell^{\textrm{th}}}
\newcommand{\first}{$1^{\textrm{st}}$}
\newcommand{\second}{$2^{\textrm{nd}}$}


\title{Extracting free-space observables from trapped interacting clusters}

%

\author{Xilin Zhang} \email{zhang.10038@osu.edu}
\affiliation{Department of Physics, The Ohio State University, Columbus, Ohio 43210, USA}
\affiliation{Physics Department, University of Washington, 
Seattle, WA 98195, USA} 

\date{April 28, 2020\\[20pt]}

\begin{abstract}
The energy spectrum of two short-range interacting particles in a harmonic potential trap 
has previously been related to free-space scattering phase shifts. 
But the existing formula for systems with a non-zero interaction range is exact only in the limit of an infinitely shallow trap. Here I provide a systematically 
improved formula---describing the low-energy dynamics---that enables the use of finite traps. 
This paves the way for extracting nuclear scattering phase shifts from {\it ab initio} nuclear many-body structure calculations, a long-sought goal in nuclear physics. 
The derivation establishes effective field theory as a powerful framework for studying the connection
between structure information of a trapped system (with two or more sub-clusters) and continuum physics in the fields of both nuclear and condensed-matter physics. 
\end{abstract}

\pacs{}

\maketitle


\paragraph{Introduction}
Nuclear experiments at low energy can not manipulate many-body systems to the extent
possible in condensed-matter or cold-atom experiments. 
However, with progress in many-body methods~\cite{Barbieri:2016uib,Barrett:2013nh,Carlson:2014vla,Lee:2008fa,Stroberg:2019mxo} and increasing computing power 
(and quantum computers~\cite{Preskill:2018}), one can start manipulating nuclear systems \emph{computationally}. 
Here, I show how trapping two clusters at low energy in a harmonic potential well tells us about
their free-space scattering through a formula connecting low-energy phase shifts 
with the confined spectrum. In this approach the 
trap compacts the system and reduces the required degrees of freedom enough to allow controlled
\textit{ab initio} calculations, as will be demonstrated elsewhere. (See e.g.,~\cite{Nollett:2006su,Navratil:2016ycn,
Elhatisari:2015iga,Shirokov:2018nlj} for other \textit{ab initio} approaches of computing light-nucleus scatterings.)

A formula for particles in a harmonic-potential trap was derived in Ref.~\cite{Busch1998}, 
and later generalized to include the full energy dependence of the phase-shift (besides the scattering length term in~\cite{Busch1998}) and for partial waves beyond s-wave~\cite{PhysRevA.65.043613,PhysRevA.65.052102,PhysRevA.66.013403,PhysRevLett.96.013201,Stetcu:2007ms,PhysRevA.80.033601,Stetcu:2010xq,Rotureau:2010uz,Blume_2012}. 
The result for angular momentum $\ell$~\cite{PhysRevA.80.033601,Stetcu:2010xq,Luu:2010hw} (called the BERW formula here) is   
\begin{eqnarray}
\!\!p^{2\ell +1 }\!\cot\delta_\ell(E)&=&(-)^{\ell+1} (4 \mr \omega)^{\ell+\half} \frac{\Gaf{\frac{3}{4}+\frac{\ell}{2} - \frac{{E}}{2\omega}}}{\Gaf{\frac{1}{4}-\frac{\ell}{2} - \frac{{E}}{2\omega}}} .  \label{eqn:master1}
\end{eqnarray}
This holds at the eigenenergies $E\equiv$ ${p^2}/2 \mr $---with the center-of-mass (CM) energy subtracted---in 
a trap where each particle experiences a potential $\omega^2 \vec{r}^2/2$ times its mass; 
$\mr$ is the reduced mass and $\delta_\ell$ the phase shift. 
Equation~\eqref{eqn:master1} is analogous to the Luscher formula~\cite{Luscher:1990ux,Beane:2003da}
that is widely applied in Lattice Quantum Chromodynamcs (for a system on a space-time torus). 

Refs.~\cite{Luu:2010hw,Rotureau:2011vf} have used \eq{eqn:master1} 
to extract nuclear scattering from  {\it ab initio} spectrum calculations. 
\begin{figure*}
  \includegraphics[width=\textwidth]{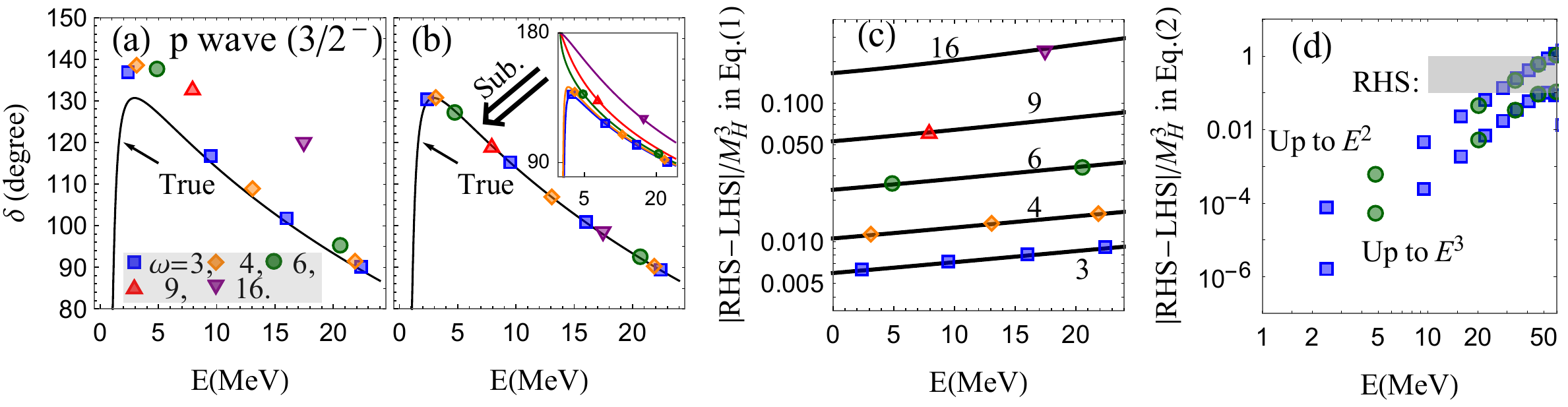}
  \caption{(a) The $n$-$\alpha$ p-wave scattering  phase shifts extracted using \eq{eqn:master1} 
  at the $\omega$-dependent eigenenergies. The ``True'' curves are the exact phase shifts. 
  (b) After subtracting
  $\omega$-dependent pieces from \textit{generalized} ERE curves (inset), the extractions 
  from \eq{eqn:master2} lie on the ``True'' curve. (c) \eq{eqn:master1}'s discrepancies at those eigenenergies. The curves plot the summation of the $C_{i\neq 0,j}$ terms in \eq{eqn:master2}'s LHS. (d) \eq{eqn:master2}'s discrepancies at those eigenenergies with two different truncations on the $i$ and $j$ indices in its LHS: ``Up to $E^2$'' and ``$E^3$''. The shaded area is the range of \eq{eqn:master2}'s RHS ($E\geq 10$ MeV). Note all the panels use (a)'s legend.  
   }\label{fig:illustration}
\end{figure*}
However, away from the infinitely-shallow-trap limit (i.e., for $\omega \neq 0$), \eq{eqn:master1} does not capture the external potential's modifications to the interaction at short distances. 
To illustrate the impact on extracting phase shifts, I use a two-body potential model~\cite{Ali:1984ds}
designed for describing neutron-$\alpha$ scattering [see the supplemental materials (SM) for details]. 
Figure~\ref{fig:illustration}(a) shows $3/2{}^-$ p-wave phase shifts extracted using \eq{eqn:master1} at the eigenenergies with $\omega=3,4,6,9,16$ MeV (typical values applicable in \textit{ab initio} calculations): they fail to align on a smooth curve and systematically deviate from the exact curve~\cite{Luu:2010hw}. 

Here, I remedy the BERW formula using pionless effective field theory (EFT)~\cite{vanKolck:1998bw,Bedaque:2002mn,Hammer:2017tjm}, which enables low-energy dynamics to be studied without specifying the
details of the short-distance physics (e.g., potential or cluster structure and excitation). 
This EFT was used to re-derive and generalize the Luscher formula~\cite{Beane:2003da,Briceno:2013lba}. 
The improved formula for a harmonic trap is   
\begin{align}
\!\!\!\!\! \sum\limits_{i,j=0}^{\infty}\! C_{i,j} {\left(\mr \omega\right)^{2i} p^{2j}}\! &= \!(\!-\!)^{\ell+1}\! (4 \mr \omega)^{\ell\!+\!\half}\!\frac{\Gaf{\frac{3}{4}\!+\!\frac{\ell}{2} \!- \!\frac{{E}}{2\omega}}}{\Gaf{\frac{1}{4}\!-\!\frac{\ell}{2}\! -\! \frac{{E}}{2\omega}}} \;, \notag \\[-2pt] 
 p^{2\ell+1}\cot\delta_\ell\left(E\right) &=   \sum_{j=0}^{\infty} C_{i=0,j} p^{2j} \;. 
\label{eqn:master2}
\end{align}
The constants $C_{i,j}$ depend implicitly on $\ell$ but are independent of $\omega$ and $p$; 
they are dimensionful and scale as proper powers of a high-momentum scale $\MH$ 
(as dictated by, e.g., the cluster excitations), 
unless there is fine tuning. When $\sqrt{\mr\omega}$ and $p$ are smaller than $\MH$, the series sum converges and thus can be truncated with a controlled error. However, outside the convergence domain, where the details of the finite-range interaction and its interplay with the trap potential are being probed, the series expansion becomes infeasible. 
It is also worth noting that in the case of {\it physically} zero-range interactions  (i.e., $\MH\rightarrow\infty$), the $C_{i\neq 0, j}$ terms, which capture the trap-induced modifications, would vanish, and \eq{eqn:master2} becomes equivalent to \eq{eqn:master1}.   

To infer the phase shifts from \eq{eqn:master2} given the eigenenergies,
the $C_{i\neq 0,j}$ terms must be simultaneously calibrated with the $C_{0,j}$. The latter determine the free-space phase shifts via the effective range expansion (ERE)~\cite{vanKolck:1998bw,Hammer:2017tjm}. 
Knowing the full potential in the n-$\alpha$ model, one can fix $C_{i,j}$ (see the SM) and
generate Fig.~\ref{fig:illustration}(b): the inset shows that the 
phase shifts extracted from \eq{eqn:master1} for a given $\omega$ sit on a curve 
parameterized by a \textit{generalized} ERE, in which the $\jth$-order coefficient is given by $\sum_{i=0}C_{i,j} (\mr\omega)^{2 i}$.  After
subtracting the trap-induced modifications, the extracted phase shifts agree with the ``True'' curve.

The essence of \eq{eqn:master2}, that the trap-induced modifications can be parameterized using a Taylor expansion in the $\omega^2$ and $p^2$ variables, can be seen in Fig.~\ref{fig:illustration}(c). The symbols show the differences between the right-hand-side (RHS) and left-hand-side (LHS) in \eq{eqn:master1}---scaled by $\MH^{-3}$---at the $\omega$-dependent eigenenergies, while the solid lines plot the summation of the $C_{i\neq 0,j}$ terms in \eq{eqn:master2}'s LHS with $ i \leq 2$ and $0\leq j \leq 3$. Indeed, they interpolate those symbols. Of course, outside the convergence domain, the series expansion would fail, as shown in Fig.~\ref{fig:illustration}(d). The differences between \eq{eqn:master2}'s two sides, based on two series truncations ($j\leq 3$ for ``Up to $E^3$'' and $j\leq 2$ for ``Up to $E^2$'', and $i\leq 2$ in both cases), are plotted against a large range of eigenenergies. The truncation errors behave as the leading terms left out of the summation in the low-energy region, but then increase to $100\%$ when the symbols reach the shaded region indicating the range of \eq{eqn:master2}'s RHS; this also suggests the breakdown scale for $E$ is between $20$ and $40$ MeV. (See the SM for more details on series convergence.)  

To extract nuclear phase shifts (or $C_{i,j}$s) from \textit{ab initio} spectra, 
\eq{eqn:master2} will play a crucial role, because \textit{ab initio} calculations, developed to computing 
compact nuclei, have uncontrolled errors when $\omega\rightarrow 0$.  
To illustrate \eq{eqn:master2}, two models are used in the SM: a hard-sphere potential model is  solved exactly, while the n-$\alpha$ model is studied numerically. 
The rest of the paper is devoted to the derivation of \eq{eqn:master2}, emphasizing 
a new set of interaction vertices between the external potential (or background field) and trapped particles, and renormalization.  

\paragraph{Derivation through EFT} 
I start by constructing an EFT Lagrangian for two spin-$0$ particles---for simplicity---in the $\lth$
partial wave, with a harmonic potential coupled to each particle. 
The framework is valid at low energies, where the details of the short-distance physics and its interplay with the trap are not resolved.  
I follow the conventions of Ref.~\cite{Zhang:2017yqc}. 
Let $c(x)$ and $n(x)$ be particle fields with masses $\mc$ and $\mn$ ($c^\ast$ and $n^\ast$ are the complex conjugations), while $\phi_{m_\ell}$ is the so-called dimer field~\cite{Kaplan:1998tg,Beane:2000fi, Bedaque:2002mn, Bedaque:2003wa, Braaten:2004rn,Briceno:2013lba,Hammer:2017tjm} with spin $l$,  projection $m_\ell$, and mass $M_{nc}=\mn+\mc$ ($\phi^{\dagger m_\ell}\equiv \left(\phi_{m_\ell}\right)^\ast$). 
The dimer $\phi$ couples to $n$-$c$ and represents the compound system.  The background field $\mypsi(x)$ is $ m  \omega^2 \vec{x}^2/2 $ in the Lab frame with $m$ as a reference mass. 
The $\phi$ propagator---and the related self-energy corrections due to $n$-$c$ multiple scattering---will be the central piece in the derivation: 
in  free space it is directly related to the $n$-$c$ scattering $T$-matrix, while in the trap its poles  give the system's spectrum. 

The Lagrangian is  $\mathcal{L}_0+\mathcal{L}_I$, where
\begin{widetext}
\begin{eqnarray}
\mathcal{L}_0 & = & \begin{pmatrix} c^\ast,\!\!& n^\ast,\!\!& \sigma_\ell\phi^{\dagger m_\ell}\end{pmatrix} \mathrm{diag}\left(i \tilde{\partial}_t + \!\frac{\vec{\partial}^2}{2 M_c}\! +\Delta_c  , i \tilde{\partial}_t + \frac{\vec{\partial}^2}{2 M_n} \!+\Delta_n , i \tilde{\partial}_t + \frac{\vec{\partial}^2}{2 M_{nc}} \! + \Delta_\ell \right) \begin{pmatrix} c,\! &   n,\! &  \phi_{m_\ell} \end{pmatrix}^T ,   \\
\mathcal{L}_{I} & = & g_\ell \phi^{\dagger m_\ell} c\left[V^{\otimes \ell}\right]_{m_\ell}\!\! n  \! + \!\mathrm{C.C} -  \phi^{\dagger m_\ell}\!\! \left[ d^{(\ell)}_{j\geq 2} \left(i \tilde{\partial}_t+ \frac{\vec{\partial}^2}{2 M_{nc}} \right)^j \!\!+ d^{(\ell)}_{j\geq0, k\geq 1} \left(i\tilde{\partial}_t+ \frac{\vec{\partial}^2}{2 M_{nc}} \right)^j\!\! \left(\frac{\mr^2}{3 m} \vec{ \partial}^2\!\mypsi\right)^k \right]\! \phi_{m_{\ell}}  . 
\end{eqnarray}
\end{widetext}
The building blocks of $\mathcal{L}_{0,I}$ are invariant under Galilean transformations, including rotation, translation, and boost. (A relevant discussion on Galilean invariance in EFT can be found e.g., in Ref.~\cite{Braaten:2015tga}.) In both lagrangians for $\psi = n$, $c$, or $\phi$, the $\psi^\ast [i\tilde{\partial}_t +\vec{\partial}^2/(2M_\psi)]\psi$ structures with $i \tilde{\partial}_t \equiv  i \partial_t - M_\psi \mypsi(x)/m$ are $\psi$'s internal energies (i.e., total energies with kinetic and external potential energies subtracted), and therefore  Galilean invariant. 

The $g_\ell$ coupling in $\mathcal{L}_I$ uses $n$-$c$'s relative velocity $\vec{V}$, while  $V^{\otimes \ell}$ denotes a rank-$\ell$ operator composed of $\ell$ copies of $\vec{V}$ 
normalized such that when $m_\ell=+\ell$, $[V^{\otimes \ell}]_{m_\ell}= [(V^{+1})^\ell]^\ast$ with $ V^{+1} \equiv -(V^x + i V^y)/\sqrt{2}$. This term means $\phi^{\dagger m_\ell}$ is coupled to a $n$-$c$ configuration having $\ell$ and $m_\ell$ as its relative angular quantum numbers. The $m_\ell$ indices are implicitly summed up so that this term is a scalar. (In general, the spin and vector indices need to be properly contracted to form scalars.) In addition, both $\vec{V}$ and spins are invariant under translation and boost, and thus the $g_\ell$ coupling preserves Galilean invariance.  Note that repeated indices in the Lagrangian (and in Eqs.~\eqref{eqn:EREforVs2}, \eqref{eq:Domega}, and \eqref{eqn:eftqc1}) are implicitly summed with specified ranges. 

It should be mentioned that the interactions in $\mathcal{L}_0$ and $\mathcal{L}_I$  with the external potential turned off follow closely previous works using a dimer-field approach~\cite{Kaplan:1998tg,Beane:2000fi,Bedaque:2002mn,Bedaque:2003wa,Braaten:2004rn,Briceno:2013lba, Hammer:2017tjm}: $\sigma_\ell$ ($=\pm1$), $\Delta_\ell$, $g_\ell$, and $d_{j}^{(\ell)}$ together reproduce the ERE (see \eq{eqn:EREforVs2} and \cite{Briceno:2013lba,Beane:2000fi}). This approach is equivalent~\cite{Kaplan:1998tg,Bedaque:2002mn,Braaten:2004rn} to the EFTs without dimer fields (see further discussion in the ``further comments'' section). 

The $d^{(\ell)}_{j\geq0, k\geq 1}$ terms are also Galilean invariant, considering the external potential $\mypsi$ is a scalar field. However, their specific structures are severely constrained by  a unique property of a harmonic potential: the CM of a multi-particle system is decoupled from its internal dynamics~\cite{Caprio:2012rv}. (It can be understood based on that the external force on the multi-particle's CM depends only on CM's location in the harmonic potential well, i.e., not affected by any other degrees of freedom.) In these couplings with ${\mr^2}\vec{\partial}^2\!\mypsi/(3 m) = \mr^2 \omega^2$, $\vec{\partial}^2$ ensures that they only shift the system's energy by $\vec{r}$-independent but $\omega^2$-dependent functions so that 
the CM behaves as a free particle in traps, i.e., decoupled from internal dynamics.

Besides powers of $\vec{\partial}^2\!\mypsi$, the other possible scalar objects built of $\mypsi$ include (1) $\mypsi^2$, $\mypsi^3$, \ldots 
(2) $(\vec{\partial}\mypsi)^4$, $(\vec{\partial}\mypsi)^6$, \ldots 
[$(\vec{\partial}\mypsi)^2$ is proportional to $\mypsi$], and (3) products of (1) and (2).  (Derivatives higher than $\vec{\partial}^2$ applied on $\mypsi$ would give zero, and therefore are not relevant here.) They would induce external potentials with powers of $\vec{r}^2$ higher than 1. Copies of $\vec{\partial}\mypsi$ can also be used to construct  tensor (vector) objects, which create anisotropic external potentials that need to be coupled to particles' momenta or spins. As the result, the new scalar and tensor objects and their products would break the CM--internal-dynamics decoupling if they are present in any interaction terms with particles.     

In principle, $\mypsi$ can be coupled to the $\phi^\ast n c$ operators (e.g., the $g_\ell$ term), which again must take the form of $\left(\vec{\partial}^2\! \mypsi\right)^{1,2,\ldots}$. However, these terms can be eliminated by rescaling the $\phi$ field by $1+ \# (\mr\omega)^2 + ... $~\cite{PhysRev.177.2239}. Since the rescaling-induced terms are already present as $d^{(\ell)}_{j,k}$ couplings in $\mathcal{L}_I$, the trap modification to $g_\ell$ is not included. 

Lastly in the free space, defining energy relative to the $n$-$c$ threshold sets $\Delta_c=\Delta_n=0$. 
Both are modified by $\mypsi$ through ``polarization'' effects as $\Delta_\ell$ by $d^{(\ell)}_{j=0,k}$ couplings, 
but they only affect the energy-references in traps and for simplicity not shown here. 

To compute the propagator of the dimer $\phi$, 
its self-energy correction due to $n$-$c$ multiple scattering needs to be included. 
A cut-off on momentum is applied to regularize loops in free space, while in traps the cut-off is applied on the virtual excitation energy~\cite{vanKolck:1998bw}. 
(However, for fine-tuned systems other schemes would be preferred, e.g., power divergence subtraction~\cite{Kaplan:1998we}.)  
Within time-independent perturbation theory~\cite{Zhang:2017yqc}, the one-loop self-energy bubble diagram in free space is 
$(2 \pi)^3 \delta(\vec{P}-\vec{P}') \delta^{m_\ell}_{m_\ell'}  \Sigma(\eL,\vec{P}) \equiv  \langle \phi_{\vec{P}'}^{m_\ell'} | H_{g_\ell} \left(\eL-H_0 + i 0^+ 
\right)^{-1} H_{g_\ell} | \phi_{\vec{P}}^{m_\ell} \rangle $. $H_0$ and $H_{g_\ell}$ are the Hamiltonians derived from $\mathcal{L}_0$ and the $g_\ell$ term in $\mathcal{L}_I$, respectively~\cite{Zhang:2017yqc}.
Both states are plane waves, with $\vec{P}$, $ \vec{P}'$,  $E_{_L}$,  $m_\ell$,  and $m_\ell'$ as $\phi$'s momenta and energy in the Lab frame, and its spin projections. (The realtionships between Feynman diagrams and the matrix elements defined here and below can be found in Ref.~\cite{Zhang:2017yqc}.) One then obtain 
\begin{align}
    & \Sigma(E) = \frac{\mathcal{A}_\ell}{\pi}  \int_{0}^{T_\Lambda} d T_q   
    \frac{(2\mr  T_q)^{\ell+\half} }{E   -T_q + i 0^+}  \notag \\ 
   & \quad\quad\ =   -\mathcal{A}_\ell \big[ip^{2\ell+1}+ \sum_{j=0}^{+\infty}L_{\ell,j}(\Lambda) p^{2j}  \big]    \;,  \notag \\ 
   &  \mathcal{A}_\ell  \equiv  
   \frac{g_\ell^2}{\mr^{2\ell-1}} \frac{2^{\ell -1} \ell!^2}{\pi(2 \ell +1)!},\;   
   L_{\ell,j}(\Lambda)\equiv  \frac{2\Lambda^{2\ell-2j+1}}{\pi\left(2\ell-2j+1\right)}. 
  \label{eqn:selfefreespace}
\end{align}
$\pe \equiv \sqrt{2 \mr \left(E+ i 0^+\right)}$, $T_q \equiv {\vec{q}^2}/(2 \mr)$, $\Lambda$ is the cut-off on  $|\vec{q}|$, and  $T_\Lambda \equiv \Lambda^2/(2 \mr)$. $E\equiv \eL-{\vec{P}^2}/(2 M_{nc})$ is the energy in the CM frame.  
Note $L_{\ell,j>\ell}(\Lambda) \rightarrow 0$ as $\Lambda\rightarrow \infty$. 

The fully dressed free-space $\phi$ propagator, which is defined through 
$ (2 \pi)^3 \delta(\vec{P}-\vec{P}')  \delta^{m_\ell}_{m'_\ell}  D(\eL,\vec{P}) \equiv \langle \phi_{\vec{P}'}^{m_\ell'}| \left[\eL-(H_0+H_I)+i 0^+\right]^{-1} | \phi_{\vec{P}}^{m_\ell} \rangle $, 
with $H_I$ from $\mathcal{L}_I$, can be computed by summing the self-energy-insertion diagrams due to $\Sigma$ and the $d_j^{(\ell)}$ vertices, yielding
\begin{eqnarray}
&& D = \frac{1}{\sigma_\ell(E+\Delta_\ell)- d^{(\ell)}_j E^j -\Sigma} 
=  \frac{-\mathcal{A}_\ell^{-1}}{\pe^{2\ell+1}[\cot\delta_\ell - i ] }, \notag \\ 
&& \text{with}\ \pe^{2\ell+1}\!\cot\delta_\ell =  \sum_{j=0}^{\infty} C_{0,j}  p^{2j}\;,\ \text{and} \notag  \\ 
&& C_{0,j} = \frac{\mathcal{A}_\ell^{-1}}{(2 \mr)^j } \!  \left\{\!-\sigma_\ell\Delta_\ell, -\sigma_\ell,  d^{(\ell)}_2\!,  d^{(\ell)}_3\!\!\ldots\right\}_{\!\!j} \!\!-L_{\ell,j}(\Lambda). \label{eqn:EREforVs2} 
\end{eqnarray}
$D$ is related to $\delta_\ell$ through the scattering $T$-matrix, which is computed by multiplying $D$ with two $g_\ell$-vertices~\cite{Zhang:2017yqc}. 
The range of the index in $d_j^{(\ell)}$ in the implicit sum is fixed in $\mathcal{L}_{I}$, 
and in the  $C_{0,j}$ definition $\{\ldots\}_j$ is the $\jth$ component of the list and $j$ is not summed. 

Now let us turn to the trapped system. Based on $\mathcal{L}_0$, one can expand $n$, $c$ and $\phi$ fields using their corresponding harmonic-oscillator wave functions~\cite{Stetcu:2010xq}. 
Again note that the $g_\ell$ coupling only picks up the $n$-$c$ configuration whose {\it total} 
angular momentum and projection equal those of the CM motion (i.e., $\phi$) and whose {\it relative}
 angular momentum and projection equal the $\phi$'s spin and projection ($\ell$ and $m_\ell$).  
 Thus the matrix element between $\phi$'s eigenstates in a trap for defining its self-energy becomes 
 $ \delta^{_{\vec{N}_\phi'}}_{_{ \vec{N}_\phi}} \delta^{m_\ell'}_{m_\ell} \Sigma_{\omega}(E) \equiv  \langle \phi_{\vec{N}_\phi'}^{m_\ell'} | H_{g_\ell} \left(\eL-H_0 \right)^{-1} H_{g_\ell} | \phi_{\vec{N}_\phi}^{m_\ell} \rangle $ 
 (note the absence of $i0^+$ in the Green's function), with 
\begin{align}   
 \Sigma_{\omega}(E) &=  \frac{g_\ell^2}{\mr^{2\ell}}  \frac{(2 \ell +1)!}{2^{\ell+2}\pi}  \sum_{n=0}^{n_\Lambda} \frac{\left(\bar{R}^{(r)}_{_{n, \ell}}(0)\right)^2 }{E- E^{(r)}_{n,\ell}} \notag \\ 
&= \frac{\mathcal{A}_\ell}{\pi}\left(4\mr\omega\right)^{\ell +\half} 
\sum_{n=0}^{n_\Lambda} f_\ell\left(\ze, n\right) \ , \notag \\ 
f_\ell\left(\ze, n\right) &\equiv \frac{ {\Gaf{n + \ell+ \threehalf}}/{\Gaf{n + 1}}}{\ze- (n +\frac{\ell}{2}+ \frac{3}{4})}.
\label{eqn:selfEomg1}
\end{align}
Here $ \ze \equiv E/(2\omega)$ and 
the relative energy $E\equiv \eL-E^{(\phi)}_{\vec{N}_\phi}$, 
with $E^{(\phi)}_{\vec{N}_\phi} = (2 N_\phi + \ell_\phi + \tfrac{3}{2}) \omega $ as the CM's energy.
If $\Delta_c$ and $\Delta_n$ receive trap-dependent ``polarization'' corrections, these corrections also need to be subtracted in defining $E$. 
In the derivation, a unitary transformation between $n$ and $c$ single-particle and CM/relative motion
eigenmodes has been used. 

Summing over the quantum numbers associated with the intermediate state's CM motion gives rise to 
the $\delta^{_{\vec{N}_\phi'}}_{_{ \vec{N}_\phi}}$ factor in defining $\Sigma_{\omega}$, 
since the CM's decoupling property is preserved and thus so is $\vec{N}_\phi$. 
For the relative dynamics, $\bar{R}^{(r)}_{_{n, \ell}}$ is part of the eigenmode function 
$R^{(r)}_{_{\vec{N}_r}}$~\cite{Stetcu:2010xq}: $R^{(r)}_{_{\vec{N}_r}}(\vec{r}) \equiv \bar{R}^{(r)}_{_{n, \ell}}( r )\, r^{\ell} Y_{\ell m_{\ell}}(\hat{r})$.  
$\vec{N}_r$ has $n,\ell$ for its radial excitation and angular momentum, 
and $E^{(r)}_{n,\ell} = (2 n + \ell + \tfrac{3}{2})\omega$.
A cut-off on $n$ is used to regularize the theory in a trap, which is in parallel with the regularization used in \eq{eqn:selfefreespace}.

The $\phi$ propagator in the trap, defined as $D_{\omega}(E)  \delta_{_{\vec{N}_\phi}}^{_{\vec{N}_\phi'}}  \delta^{m_\ell'}_{m_\ell}$, can be computed by summing up all self-energy insertion diagrams, including insertions of $\Sigma_\omega$ and those of the $d_j^{(\ell)}$ and $d_{j,k}^{(\ell)}$ vertices. One get 
\begin{widetext}
\begin{eqnarray}
&& D_{\omega} = \frac{1}{\sigma_\ell(E+\Delta_\ell)- d^{(\ell)}_j E^j -\Sigma_{\omega}(E)- d^{(\ell)}_{j, k} E^j (\mr\omega)^{2k}  } =  \frac{(-) \mathcal{A}_\ell^{-1}}{\pe^{2\ell+1}\cot\delta_\ell \!+\! \tfrac{1}{\mathcal{A}_\ell}\!\left[\Sigma_{\omega}(E) \!-\! \mathcal{P}\Sigma(E)  \!+\!   d^{(\ell)}_{j, k} E^j (\mr \omega)^{2k} \right]}. \label{eq:Domega} 
\end{eqnarray} 
\end{widetext}
In the 2nd step, the principal value of the free-space self-energy $\mathcal{P}\Sigma$ is added and subtracted. 
Thus, the quantization condition can be derived by setting the denominator in \eq{eq:Domega} to zero:  
\begin{eqnarray}
\!\!\!\pe^{2\ell+1}\!\cot\delta_\ell(E)\! +\!  \frac{d^{(\ell)}_{j, k}}{\mathcal{A}_\ell} E^j (\mr\omega)^{2k} \!=\! \frac{\mathcal{P}\Sigma(E)\!-\!\Sigma_{\omega}(E)}{\mathcal{A}_\ell}. \label{eqn:eftqc1}
\end{eqnarray}
There exists a special relation between $\Lambda$ (or $\bar{T}_\Lambda \equiv T_\Lambda/\omega$) and $n_\Lambda$ such that the divergences in $\Sigma$ and $\Sigma_\omega$ cancel in \eq{eqn:eftqc1}, and thus $d^{(\ell)}_{j,k}$ are finite. For s-wave, $\bar{T}_\Lambda = 2 n_\Lambda (1+ O(n_\Lambda^{-1}))$, but for p-wave the $n_\Lambda^{-1}$-order term needs to be specified: ${\bar{T}_{\Lambda}} = 2 n_\Lambda \left(1+\frac{7}{4}  n_\Lambda^{-1} + O(n_\Lambda^{-2})\right)$; for d-wave  another higher order term needs to be specified: $\bar{T}_{\Lambda} = 2 n_\Lambda \left(1+\frac{9}{4}  n_\Lambda^{-1}- \frac{37}{32}  n_\Lambda^{-2} + O(n_\Lambda^{-3})\right)$; for even larger $\ell$, more terms need to be specified accordingly. 
Details on the renormalization can be found in the SM. However, the above $n_\Lambda$-$\Lambda$ relations should be considered as a specific scheme; any alternative ones  would need to ensure that the divergences in  \eq{eqn:eftqc1}'s right side can be absorbed by the $d^{(l)}_{j,k}$ terms in the left side so that phase shifts are cut-off independent and the CM-decoupling property is not violated. 

The right side of \eq{eqn:eftqc1} in this scheme then becomes    
\begin{eqnarray}
 && -\frac{1}{\pi} \left(4\mr\omega\right)^{\ell+\half} \bigg[\sum_{n=0}^{n_{\Lambda}}f_\ell\left(\ze, n\right)  + \pi
\sum_{j=0}^{\ell}\ze^j L_{l,j}\!\!\left(\!\!\sqrt{\tfrac{\bar{T}_\Lambda}{2}}\right)  \bigg] \notag \\ 
 && \equiv    -\frac{1}{\pi} \left(4\mr\omega\right)^{\ell+\half} \sum_{n=0}^{(\mathcal{R})} f_\ell\left(\ze,n\right),   \label{eqn:regl1}
\end{eqnarray} 
with ``$(\mathcal{R})$'' labeling the renormalized series sum with $n_\Lambda \rightarrow +\infty$. To finish the derivation, this identity is needed:  
\begin{eqnarray}
 \sum_{n=0}^{(\mathcal{R})} f_\ell(z, n)  = (-)^\ell \pi \,  \frac{\Gaf{\frac{\ell}{2}\!+\! \frac{3}{4} \!-\! z}}{\Gaf{\frac{1}{4}\!-\!\frac{\ell}{2}\! -\!z}} , \label{eqn:conj2}
\end{eqnarray} 
which holds in the entire complex $z$ plane 
(both sides have the same poles and residues, see the proof in the SM). 
By redefining  $d^{(\ell)}_{j,i}\equiv \mathcal{A}_\ell(2\mr)^j  C_{i,j}$ in \eq{eqn:eftqc1} and applying \eq{eqn:conj2} in \eq{eqn:regl1}, \eq{eqn:eftqc1} gives  \eq{eqn:master2}. 

\paragraph{Further comments} 
It is worth comparing $D(E)$ in \eq{eqn:EREforVs2} and $D_\omega(E)$ in \eq{eq:Domega} in the 
complex $E$ plane. 
 $1/D(E)$ has a branch cut---known as the unitary cut---on the positive real axis due to the $-ip^{2\ell+1}$ term, which changes into a series of poles---called ``unitary'' poles below---for  $1/D_\omega(E)$ (from the term $[\Sigma_{\omega}(E)-\mathcal{P}\Sigma(E)]/\mathcal{A}_\ell$).
 Both non-analyticities are directly connected to unitarity and thus independent of framework, power counting, and fine tuning. 
 
 However, fine tuning and power counting do impact the behavior of the ERE function~\cite{vanKolck:1998bw}: in a natural case, $C_{0,j} \sim M_{_{H}}^{2\ell+1-2j} $; in a \first\ fine-tuned case, $C_{0,0}$ is enhanced;  
 and in a \second\ fine-tuned case, the function has low-energy poles. Note Ref.~\cite{vanKolck:1998bw} uses EFTs {\it without} a dimer field; and the Lagrangian is  the same for the three cases, except power countings. One can add the couplings between $\mypsi(x)$ and particles to the Lagrangian, again by multiplying the short-distance-interaction terms with powers of $\vec{\partial}^2\mypsi$, e.g., for s-wave $\mathcal{L}_I= [\tilde{d}_0 + \sum_{k\geq 1}\tilde{d}_{0, k} (\mr^2 \vec{ \partial}^2\!\mypsi/3m )^k  ] (c\, n)^\ast (c\, n) + \dots$.  
 
For each of these cases, the $T$-Matrix in a trap can be computed in the same way as the free-space one~\cite{vanKolck:1998bw} but with the bare couplings substituted by the corresponding modified ones---e.g., $\tilde{d}_0 \rightarrow \tilde{d}_0+\sum_{k\geq 1}\tilde{d}_{0, k}(\mr \omega)^{2k}$---and the unitary cut by the ``unitary'' poles. Since the EFT calculations reproduce the free-space $T$-matrix using ERE parameters $C_{i=0,j}$, the $T$-matrix in a trap can be parameterized in the same way but with $\omega$-dependent ERE parameters. The relation between these ERE parameters and the bare couplings is nonlinear, but the former's $\omega$-dependence could be expanded in terms of $\omega^2$. (This expansion must be examined with care, if its convergence radius is much smaller than the naive estimate based on $\MH$, e.g., due to fine-tuning of the trap's modification to the interaction at short distance.)  Finally, by identifying the poles of the trap $T$-Matrix, one then reproduces \eq{eqn:master2} for the natural and the \first\ fine-tuned case; for the \second\ fine-tuned case, a Laurent expansion of $p^{2\ell+1}\cot\delta_\ell$ was derived~\cite{vanKolck:1998bw}, so the same expansion should be used in \eq{eqn:master2}'s left side with the parameters carrying $\omega^2$ corrections. In other words in my approach {\it with} a dimer field, resumming of $d_j^{(\ell)}$ and $d_{j,k}^{(\ell)}$ terms is needed.

\paragraph{Summary}
I have applied pionless EFT to two short-range interacting particles 
in an external harmonic trap to derive 
a systematically improved BERW formula that is exact even 
at finite $\omega$. 
It is valid when the infrared scale of the trap ($\sqrt{\mr \omega}$) 
and the relative momentum ($p$) are both smaller than the high momentum scale set by the dynamics. 
This provides a firm foundation for implementing a Luscher-formula-like approach to connect 
nuclear scattering and {\it ab initio} structure calculations. 
The derivation involved new coupling terms between the background field and particles, which 
lead to the improvements of the original BERW formula. 
Moreover, a careful analysis of renormalization shows a non-trivial relation 
between the cut-off $\Lambda$ on relative momentum in free space and cut-off $n_\Lambda$ on 
the number of radial excitation in a trap. 
The renormalization procedure is further confirmed by \eq{eqn:conj2}'s proof. 
Both aspects 
are instructive for deducing connections between a trapped system (with two or more clusters) 
and free-space scattering/reactions for both nuclear and cold atom physics~\cite{Blume_2012}. It should also be interesting to apply this framework to study exotic atoms\footnote{Here the long-range interaction is the attractive Coulomb force. The so-called  Deser–Trueman formula~\cite{Combescure:2007ki} relates exotic atom's energy levels to the short-distance interaction's scattering length.} and quantum dots~\cite{Combescure:2007ki}.  

\paragraph{Acknowledgment}  
I would like to thank Dick Furnstahl, Chan Gwak, Jason Holt, David Kaplan, Ubirajara Van Kolck, Petr Navratil, Daniel Phillips, Martin Savage, Ragnar Stroberg, and Chieh-Jen Yang for helpful discussions, and Yuri Kovchegov for pointing out how to use contour integration to prove \eq{eqn:conj2}. I also thank Dick Furnstahl and Jordan Melendez for careful proofreading of the manuscript. The work was supported by the National Science Foundation under Grant No. PHY–1614460 and the NUCLEI SciDAC Collaboration under US Department of Energy MSU subcontract RC107839-OSU, the US Department of Energy under contract DE-FG02-97ER-41014, and the US Institute for Nuclear Theory.

\bibliographystyle{apsrev4-1}
\bibliography{RevisitBERWformula_PRCRapid_2ndSub_Main_SM_for_arXiv}

\clearpage
\onecolumngrid
\begin{center}
  \textbf{\large Supplementary Material for Extracting free-space observables from trapped interacting clusters}\\[.4cm]
  Xilin Zhang\\[.1cm]
  {\itshape Department of Physics, The Ohio State University, Columbus, Ohio 43210, USA} \\
  {\itshape Physics Department, University of Washington, Seattle, WA 98195, USA} 
\end{center}

\twocolumngrid

\setcounter{equation}{0}
\setcounter{figure}{0}
\setcounter{table}{0}
\setcounter{section}{0}
\setcounter{page}{1}
\makeatletter
\renewcommand{\theequation}{S\arabic{equation}}
\renewcommand{\thefigure}{S\arabic{figure}}

\maketitle

\section{An exactly solvable case: hard-sphere potential}
I demonstrate here that if the interaction is short-ranged and has the form of a hard sphere, the parameters in the improved BERW formula (i.e., Eq.~(2) in the main text) can be found analytically. 
This model was studied in Ref.~\cite{PhysRevA.65.052102} by using parabolic cylinder functions for s-wave channel. 
The hard-sphere potential is defined as $V_s(r)=  +\infty $ if $r \leq r_c $ and $0$ otherwise ($r\equiv |\vec{r}|$ with $\vec{r}$ the relative displacement between the two particles). 
In addition, each particle experiences an external harmonic potential. 
Because the CM motion is factorized, one can just focus on the relative motion; the corresponding external potential is $\mr \omega^2 \vec{r}^2/2$, with $\mr$  the reduced mass. 

Let us define $\bar{r}\equiv r/b$ and $\bar{r}_c\equiv r_c/b$ with $b\equiv 1/\sqrt{\mr \omega}$, and $\bar{E} \equiv E/\omega$. 
When $\bar{r}>\bar{r}_c$, the Schr\"odinger equation in the $\ell^{\textrm{th}}$ partial wave becomes
\begin{eqnarray}
\left[ -\frac{d^2}{ d \bar{r}^2} + \frac{\ell(\ell+1)}{\bar{r}^2} + \bar{r}^2 \right] u_\ell = 2 \bar{E} u_\ell  \; , 
\end{eqnarray} 
with the radial wave function defined as $u_\ell(\bar{r})/r$. Thus, the wave function at $\bar{r}>\bar{r}_c$ is a linear combination of two independent solutions to the harmonic oscillator Schr\"odinger equation~\cite{PhysRevA.80.033601}: 
\begin{eqnarray}
u_\ell & = & e^{- \frac{\bar{r}^2}{2}} \bigg[c_1 \bar{r}^{\ell+1} M\left(\frac{\ell}{2} + \frac{3}{4} -\frac{ \bar{E}}{2}, \ell+\frac{3}{2} , \bar{r}^2\right) 
\notag \\ 
&& + c_2 \bar{r}^{-\ell} M\left(-\frac{ \ell}{2} + \frac{1}{4} -\frac{ \bar{E}}{2}, - \ell +\frac{1}{2}, \bar{r}^2\right)\bigg]  \; , 
\end{eqnarray}
where $M(a,b,z)$ is the Kummer function~\cite{MathHandBook1}. 

When $\bar{r} \rightarrow +\infty$, $u_\ell$ should go to zero to guarantee that the wave function is normalizable. Since at large $z$ \cite[Eq.~13.7.2]{NIST:DLMF}
\begin{eqnarray}
\!\!\!\!\!\!\! M\left(a, b, z\right) \approx \Gaf{b}\! \left[ \frac{z^{a-b} e^z}{\Gaf{a}} \! + \! \frac{(-z)^{-a}}{\Gaf{b-a}} \right]\!\!\left[1\!+\!O(z^{-1})\right],  
\end{eqnarray}
the following is required: 
\begin{eqnarray}
c_1  \frac{\Gaf{\ell+\frac{3}{2}} }{\Gaf{\frac{\ell}{2} +\frac{3}{4} - \frac{\bar{E}}{2}}}  + c_2 \frac{\Gaf{-\ell+\frac{1}{2}} }{\Gaf{-\frac{\ell}{2} +\frac{1}{4} - \frac{\bar{E}}{2}}}  = 0 \ . \label{eqn:hsquantization1}
\end{eqnarray} 
Meanwhile, $u_\ell \left(\bar{r}_c\right)=0$ implies 
\begin{align}
c_1 \bar{r}_c^{\ell+1} M&\left(\frac{\ell}{2} + \frac{3}{4} -\frac{ \bar{E}}{2},  \ell+\frac{3}{2} , \bar{r}_c^2\right) \notag \\ 
+&\frac{c_2}{\bar{r}_c^\ell} M\left(-\frac{ \ell}{2} + \frac{1}{4} -\frac{ \bar{E}}{2}, - \ell +\frac{1}{2}, \bar{r}_c^2\right) = 0 \ . \label{eqn:hsquantization2}
\end{align}
\eqtoeqp{eqn:hsquantization1}{eqn:hsquantization2} have nontrivial solutions only if 
the corresponding determinant is zero, which gives the quantization condition:
\begin{eqnarray}
 \!\! \!\!\!\frac{\left(2\bar{r}_c\right)^{2\ell+1}}{(-)^\ell\mathcal{N}_\ell}\frac{\Gaf{\frac{\ell}{2} +\frac{3}{4} - \frac{\bar{E}}{2}}}{\Gaf{\frac{1}{4}-\frac{\ell}{2} - \frac{\bar{E}}{2}}} \!=\! \frac{ M\!\left(\frac{1}{4}\!-\!\frac{ \ell}{2}\! -\!\frac{ \bar{E}}{2}, \frac{1}{2} \!-\! \ell , \bar{r}_c^2\right)}{M\!\left(\frac{\ell}{2}\! +\! \frac{3}{4} \!-\!\frac{ \bar{E}}{2}, \ell\!+\!\frac{3}{2} , \bar{r}_c^2\right)}. \  \label{eqn:hardsphereQC}
\end{eqnarray}
Here, $\mathcal{N}_\ell \equiv (2 \ell+1 )!! (2 \ell -1)!! $ [$(-1)!! \equiv 1 $]. 

Note that the left side of \eq{eqn:hardsphereQC} is the right side of the main text's Eq.~(1) 
 multiplied by a  
$(-){r_c^{2\ell +1}}/{\mathcal{N}_\ell}$ factor. Meanwhile, the phase shift due to $V_s(r)$ 
in the $\ell^{\textrm{th}}$ partial wave is $\tan\delta_\ell = {j_\ell(p r_c)}/{y_\ell (pr_c)}$~\cite{JoachainQCT1975}. 
Thus the discrepancy of the BERW formula is identified as the difference between the left side of 
the main text's Eq.~(1)  multiplied by $ - {r_c^{2\ell +1}}/{\mathcal{N}_\ell}$ 
and the right side of \eq{eqn:hardsphereQC},  i.e., 
\begin{eqnarray}
\frac{\left( p  r _c\right)^{2\ell +1} }{(-)\mathcal{N}_\ell}
 \frac{y_\ell(p  r _c)}{j_\ell p  r _c)}\, \overset{?}{=}\, \frac{ M\left(\frac{1}{4}-\frac{ \ell}{2}  -\frac{ \bar{E}}{2}, \frac{1}{2}- \ell, \bar{r}_c^2\right)}{M\left(\frac{\ell}{2} + \frac{3}{4} -\frac{ \bar{E}}{2}, \ell+\frac{3}{2} , \bar{r}_c^2\right)}  \ . \label{eqn:test1}
\end{eqnarray}
 The left side can be expanded in terms of $\left(p r_c\right)^2$: 
\begin{eqnarray}
\!\!\!\!\! 1\!+\! \frac{(2 \ell +1) \left(p r_c\right)^2}{4 \ell (\ell+1)-3 } \!+\! \frac{(\ell+3)(2\ell+1) \left(p r_c\right)^4 }{(2 \ell -3)(2\ell+3)^2 (2\ell+5)} 
 \!+\! \cdots  \label{eqn:hsere} 
\end{eqnarray} 
Now according to Ref.~\cite[Eq.(13.2.2)]{NIST:DLMF}, $M(a,b,z)$ can be expanded as 
$1+\frac{a}{b} z +\frac{a(a+1)}{b(b+1)}\frac{z^2}{2!}+\dots $, indicating that 
\eq{eqn:test1}'s right side, a function of $E$ and $r_c$, can be approximated by a double expansion 
in terms of powers of $\bar{r}_c^2$ and $2 \bar{E}\bar{r}_c^2 = (pr_c)^2$ when both are small
(note that the coefficient denominators in the expansion of the $M$ functions in \eq{eqn:test1} are independent of $\bar{E}$ and $\bar{r}_c$). 
This also suggests the difference between left and right sides in \eq{eqn:test1} can be expanded 
in terms of $\bar{r}_c^2$ and $2 \bar{E}\bar{r}_c^2$. 

It can be shown that with $\omega \to 0$ and $E$ and $r_c$ fixed, the right side of \eq{eqn:test1} approaches its left side. In this limit, $\bar{E} \to \infty$,   
\begin{align}
    \lim_{\omega\rightarrow 0} M &\left(-\frac{ \ell}{2} 
    + \frac{1}{4} -\frac{ \bar{E}}{2}, - \ell +\frac{1}{2}, \bar{r}_c^2\right)\notag \\ 
   &=\lim_{\omega\rightarrow 0} M\left(-\frac{E}{2\omega}, 
   - \ell +\frac{1}{2}, r_c^2 \mr \omega \right) \notag \\ 
&= {}_0F_1\left(; -\ell+\half; -\half \mr E r_c^2\right) \; ,  \\
  \lim_{\omega\rightarrow 0} M &\left(\frac{\ell}{2} + \frac{3}{4} -\frac{ \bar{E}}{2}, \ell+\frac{3}{2} ; \bar{r}_c^2\right) \notag \\ 
 &= {}_0F_1\left(; \ell+\frac{3}{2}; -\half \mr E r_c^2\right) \; , 
\end{align}
where ${}_0F_1$ is a Confluent Hypergeometric Limit Function.  Reference~\cite{Petkovsek1996} suggests
\begin{eqnarray}
J_\alpha(z)=\frac{\left(\frac{z}{2}\right)^\alpha}{\Gamma\left(\alpha+1\right)} {}_0F_1\left(;\alpha+1;-\frac{z^2}{4}\right)  \; . 
\end{eqnarray}
Since 
\begin{eqnarray}
\!\!\!\!\! j_\ell(z) \!=\! \sqrt{\frac{\pi}{2z}} J_{\ell+\frac{1}{2}} (z), \  y_\ell(z) \!=\! (\!-\!)^{\ell+1} \!\!\sqrt{\frac{\pi}{2z}} J_{-\ell-\frac{1}{2}} \!(z) \; , 
\end{eqnarray}
one then obtain 
\begin{align}
\frac{{}_0F_1\left(; -\ell+\half; -\half \mr E r_c^2\right)}{{}_0F_1\left(; \ell+\frac{3}{2}; -\half \mr E r_c^2\right) } & = \frac{\left( p  r _c\right)^{2\ell +1} }{(-)\mathcal{N}_\ell}  \frac{y_\ell(p  r _c)}{j_\ell (p  r _c)} \  ,   
\end{align}
and thus prove that the difference between the two sides of \eq{eqn:test1} disappears as $\omega\rightarrow 0$. 
This requires that the difference, if expanded in terms of $\bar{r}_c^2$ and $2 \bar{E}\bar{r}_c^2$, must have positive powers of $\bar{r}_c^2$ ($\propto \omega$).

 Moreover, both sides of \eq{eqn:test1} are unchanged when 
 $\bar{E}\to - \bar{E}$ and $\bar{r}_c^2 \to -\bar{r}_c^2$ (i.e., $\omega \to -\omega$), 
 because $M(a, b, z)= e^z\, M(b-a,b,-z)$\cite[Eq.(13.2.39)]{NIST:DLMF}, 
 indicating that the powers of $\bar{r}_c^2$ are positive and even 
 (the powers of $2 \bar{E}\bar{r}_c^2$ are non-negative integers). 
 Therefore, the following expansion is expected, with the first two terms explicitly given: 
\begin{align}
  \frac{(-1)^{\ell+1}}{\mathcal{N}_\ell} \Bigl[&\frac{2 r_c}{b}\Bigr]^{2 \ell+1} 
    \frac{\Gamma\left(\frac{3}{4}+\frac{\ell}{2} - \frac{\bar{E}}{2}\right)}{\Gamma\left(\frac{1}{4}-\frac{\ell}{2} - \frac{\bar{E}}{2}\right)} 
  -\frac{\left(p r_c\right)^{2\ell +1 } \cot\delta_\ell\left(p\right)}{\mathcal{N}_\ell}   \notag \\ 
& =  \frac{(2 \ell +1) \left(\frac{r_c}{b}\right)^4}{2 (2\ell-3) (2\ell +5)} \notag \\
  & \quad\null +\frac{(2 \ell +1)  (6 \ell +25)  (p r_c)^2 \left(\frac{r_c}{b}\right)^4 }{6 (2 \ell -5) (2 \ell +3) (2 \ell +5) (2 \ell +7) }  + \dots  \notag  \\ 
& \equiv   \sum\limits_{i=1}^{\infty} \sum_{ j=0}^{\infty} \overline{C}_{i,j} \left(\frac{r_c}{b}\right)^{4i}\left( pr_c\right)^{2j} \; . \label{eqn:gEREforHSmodel}
\end{align}
Meanwhile, the ERE for $V_s$ can be derived easily from \eq{eqn:hsere}: $\left({p r_c }\right)^{2\ell+1}  \cot\delta_\ell\left(p\right)/\mathcal{N}_\ell =  \sum_{j=0}^{\infty} \overline{C}_{i=0,j} \left({p r_c}\right)^{2j}$. If a high-momentum scale $M_{_{H}} = r_c^{-1}$ is identified, \eq{eqn:gEREforHSmodel} can be transformed to the improved BERW formula after redefining $C_{i,j}= \mathcal{N}_\ell M_{_{H}}^{2\ell+1-4i-2j} \overline{C}_{i,j}$. 

\section{The 2nd model for numerical testing}

\begin{figure}
\includegraphics[width=0.4 \textwidth]{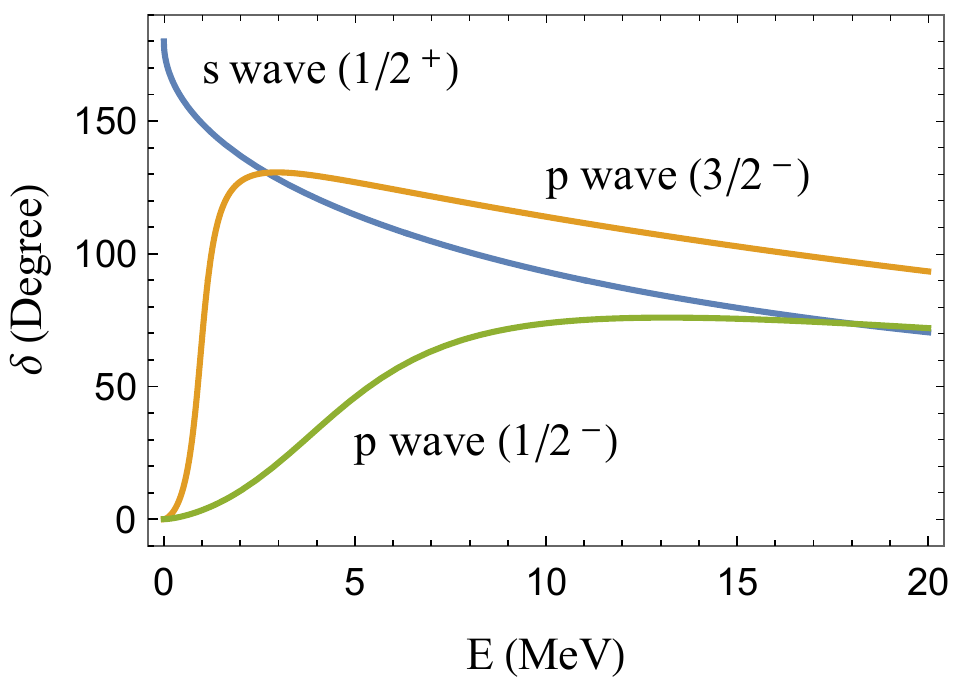}
\caption{The neutron-$\alpha$ s-wave and p-wave scattering phase shifts vs.\ their CM energy $E$, as produced by a model square-well potential (see text)~\cite{Ali:1984ds}.}\label{fig:modelps}
\end{figure}

\begin{figure*}
\includegraphics[width=0.325\textwidth]{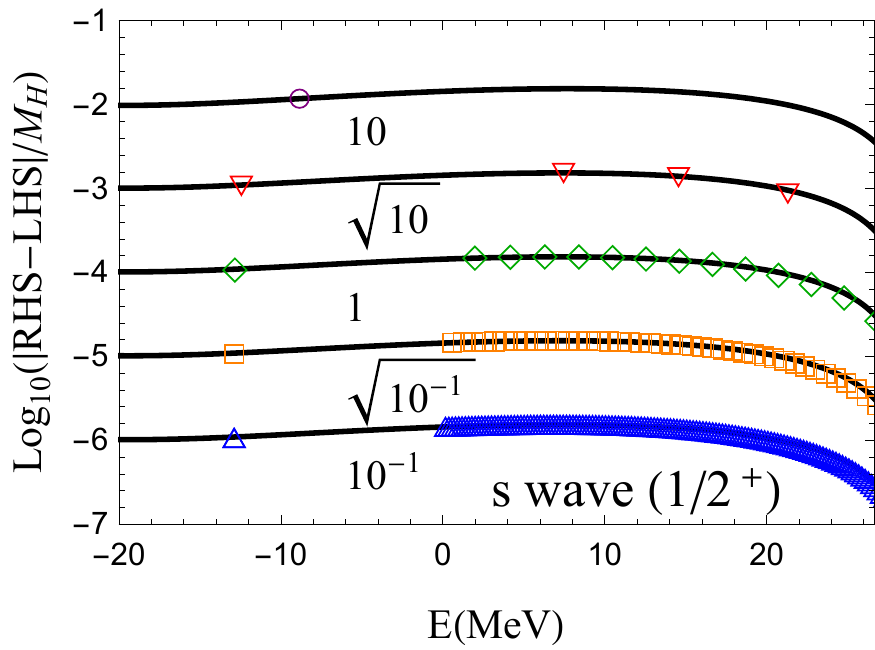}
\includegraphics[width=0.325\textwidth]{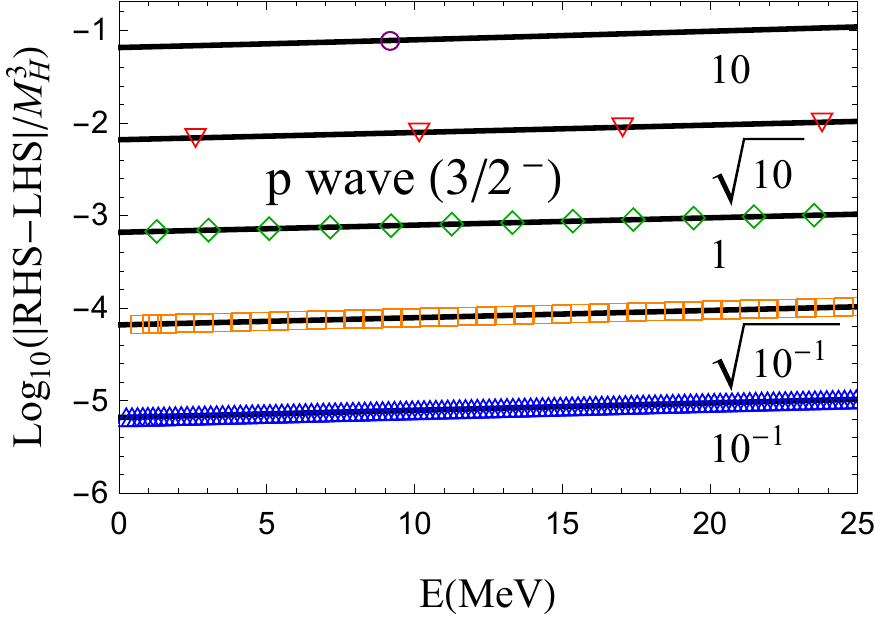}
\includegraphics[width=0.325\textwidth]{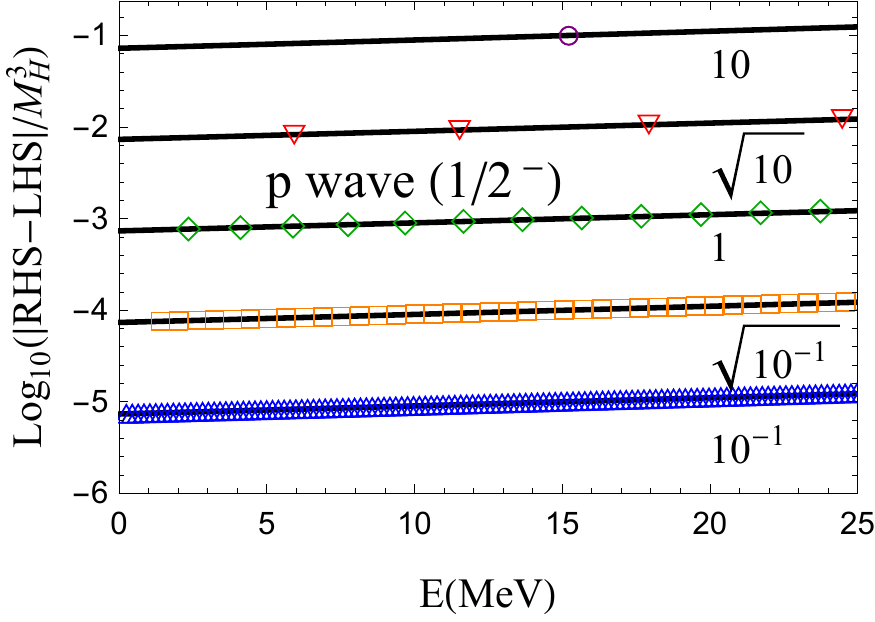}
\caption{The $y$-axis is $\log_{10}\left(|\mathrm{RHS}-\mathrm{LHS}|/\MH^{2\ell+1}\right)$ 
(where RHS/LHS means right-/left-hand-side of the equation) 
in the original BERW formula 
(Eq.~(1) in the main text) with 
$\MH=200 \mathrm{MeV}$, while the $x$-axis is the energy. 
Three different channels are plotted in different panels. The symbols are the differences computed 
at those exact eigenenergies corresponding to harmonic potential traps with six $\omega$ values: 
$0.1$, $\sqrt{0.1}$, $1$, $\sqrt{10}$ and $10$ MeV. Meanwhile, the curves are the trap-dependent 
corrections in the improved-BERW formula, with $C_{i\neq0,j}$ parameters fitted as described 
in the text. }\label{fig:check}
\end{figure*}

In contrast to the analytical study in the last section, here a simple square-well model~\cite{Ali:1984ds} 
is used to numerically demonstrate that the discrepancy of the BERW formula can be expanded in terms of powers of $\omega^{2}$ and $p^2$, 
 which is the essence of the improved BERW formula. The potential was constructed to qualitatively describe neutron-$\alpha$ scattering in the s- and p-waves~\cite{Ali:1984ds}. 
The quantum numbers for the considered channels in $J^\pi$ notation are ${\frac{1}{2}}^+$, ${\frac{1}{2}}^-$ and ${\frac{3}{2}}^-$.  
The potential $V_s(r) =  V_0 (1+ \beta \vec{L}\cdot\vec{\sigma})$ when $r < r_c$ and $0$ when $r > r_c$, with $V_0=-33\,$MeV, $r_c=2.55\,$fm, $\beta=0.103$~\cite{Ali:1984ds}. 
$\vec{L}\cdot \vec{\sigma}$ is the spin-orbit coupling, 
which generates differences between the $p_{\frac{3}{2}}$ and  $p_{\frac{1}{2}}$ channels. 

In this exercise, the potential is treated as the exact underlying physics for the two particles. 
The phase shifts, calculated by solving the corresponding Schr\"odinger equations in the continuum, are shown in Fig.~\ref{fig:modelps}.  
These phase shifts are considered to be ``exact'' ones. 
(The ${\frac{3}{2}}^-$ channel's exact phase shift is also shown in Fig.~1 in the main text.) 
Meanwhile, the spectra for the two particles in various harmonic potential traps can also be precisely computed. 
The goal is to test the original and the improved BERW formulas using these exact phase shifts and energy spectra.

\begin{table}
\begin{ruledtabular} 
   \begin{tabular}{cccc}
    & $i=0$ & $1$ & $2$        \\ \hline  
$j=0$ &  $-$0.3536 & $-$0.4097 & 0.004206 \\
$1$ &  0.7898 & $-$0.1756 & $-$0.4063  \\
$2$ &  0.1743 & 0.2010 &   \\
$3$ & $-$0.01278 & 0.2949  &   \\
\end{tabular} \caption{$\overline{C}_{i,j}$ for the $s_{1/2}$ channel.} \label{tab:s1/2}
	\end{ruledtabular}
	\end{table}

\begin{table}
\begin{ruledtabular} 
   \begin{tabular}{cccc}
    & $i=0$ & $1$ & $2$        \\ \hline  
$j=0$ &  0.02304 & $-$1.870 & 0.5400 \\
$1$ &  $-$0.5571 & $-$0.8676 & 0.3118 \\
$2$ &  0.5609 & $-$0.2829 &    \\
$3$ & 0.08738 & $-$0.03911  &  \\
\end{tabular} \caption{$\overline{C}_{i,j}$ for the $p_{3/2}$ channel. } \label{tab:p3/2}
	\end{ruledtabular}
	\end{table}

	\begin{table}[!h]	
\begin{ruledtabular} 
   \begin{tabular}{cccc}
    & $i=0$ & $1$ & $2$        \\ \hline  
$j=0$ & 0.1390 & $-$2.084 & 0.7400 \\
$1$ & $-$0.4427 & $-$1.090 & 0.5210 \\
$2$ & 0.6225 & $-$0.4064 &   \\
$3$ & 0.1114 & $-$0.08700  & \\
\end{tabular} \caption{$\overline{C}_{i,j}$ for the $p_{1/2}$ channel. } \label{tab:p1/2}
	\end{ruledtabular}
	\end{table}

In Fig.~\ref{fig:check}, the discrete symbols are the differences between the two sides in the main text's
Eq.~(1) 
divided by $\MH^{2\ell+1}$ at eigenenergies associated with the trap frequency 
$\omega=0.1$, $\sqrt{0.1}$, $1$, $\sqrt{10}$ and $10$\,MeV. 
Here, I take $\MH = 200$\,MeV ($\sim 1\;\mathrm{fm}^{-1}$), as motivated by the value of $r_c$. One can see that the leading-order difference does scale as $\omega^2$. 
Also, in the s-wave channel there is a deep bound state in free space---unphysical for the n-$\alpha$ system---and thus a distinct negative eigenenergy in all the traps. 
However, for the other two channels no bound state exists in free space.  

To see how well the improved BERW formula works, one need to know the values of $C_{i,j}$ corresponding to this particular potential. 
First, {\it another} set of eigenenergies at extremely small trap frequencies 
(both on the order of $10^{-6}$ MeV) is computed. Second, they are used as inputs to fit $C_{i,j}$ values
based on the improved BERW formula. A least-squares fit uses as the objective function the sum 
of the squares of the differences between the two sides in the improved BERW formula, 
 as calculated at those 
small eigenenergies. The small values of $\omega$ and eigenenergies help separate the impact of 
$C_{i,j}$ at different orders, and enables a precise fit (it amounts to computing derivatives at 
$E=0$ and $\omega=0$ with points separated by tiny distances).  The series in the improved BERW formula 
is nonetheless truncated to $i\leq 2$ and $j\leq 3$. 
To make the parameters dimensionless, one can rescale them by appropriate powers of $\MH$,  $C_{i,j} \equiv \overline{C}_{i,j} \MH^{2\ell+1-4i-2j}$. The best-fit values for $\overline{C}_{i,j}$ are shown in Tables~\ref{tab:s1/2}, \ref{tab:p3/2}, and~\ref{tab:p1/2}.

 A few higher-order $\bar{C}_{i,j=2}$ values are not shown in those tables, because 
 over-fitting~\cite{Furnstahl:2014xsa} in this simple exercise leads to values significantly 
 larger than $1$. However, these contributions are very small in the plots shown if they on the order 
 of $1$ (their natural size), and therefore they are set to zero in generating the plots. 

Having determined the $C_{i,j}$, the $C_{i\neq0,j}$ terms in the improved BERW formula 
are used in Fig.~\ref{fig:check} to interpolate between the discrete symbols, 
i.e., the difference between the right and 
left sides of the BERW formula 
as computed at discrete eigenenergies up to $\approx 20$ MeV. 
Because the fitting of $C_{i,j}$ is carried for $E$ and $\omega$ near zero ($\sim 10^{-6}$ MeV),
the agreement with the interpolating curves and discrete symbols demonstrates that the discrepancy 
of the BERW formula 
can indeed be expanded in terms of powers of $\omega^{2}$ and $p^2$, 
as implied by the improved BERW formula.

\begin{figure*}
\includegraphics[width=0.325\textwidth]{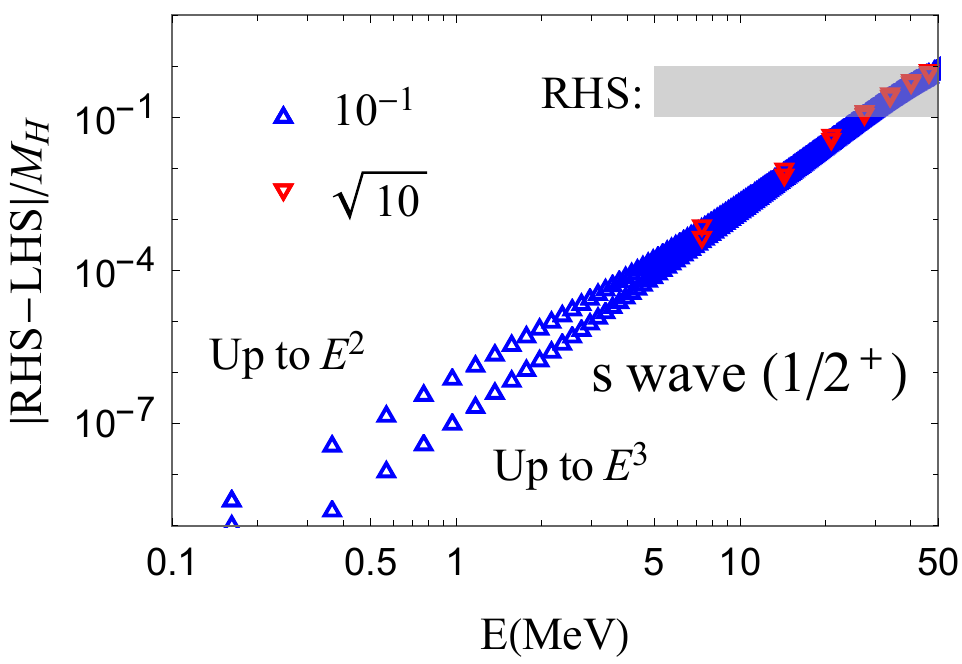}
\includegraphics[width=0.325\textwidth]{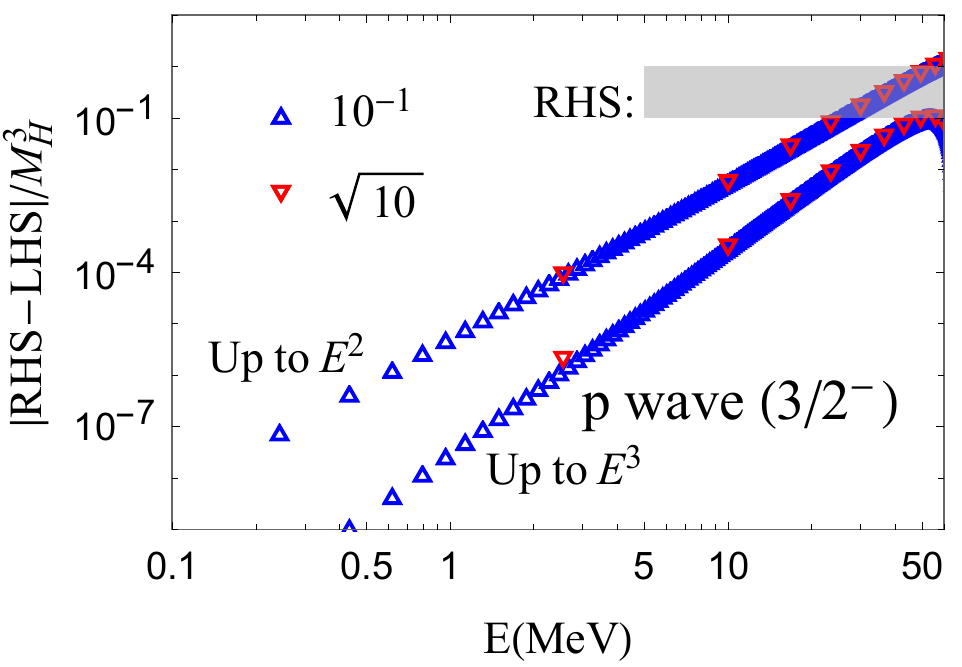}
\includegraphics[width=0.325\textwidth]{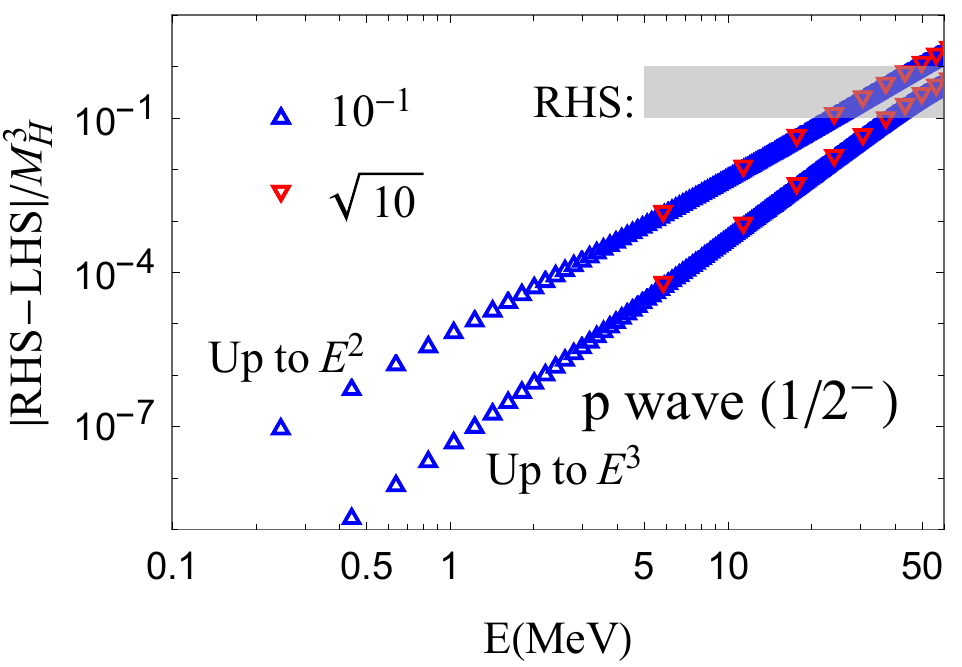}
\caption{The $y$-axis is $\log_{10}\left(|\mathrm{RHS}-\mathrm{LHS}|/\MH^{2\ell+1}\right)$ in the {\it improved} BERW formula 
(Eq.~(2) in the main text) with $\MH=200 \mathrm{MeV}$, while the $x$-axis is the energy.  Three different channels are plotted in different panels. The symbols are the differences computed  at those exact eigenenergies corresponding to harmonic potential traps with $\omega=0.1$ and $\sqrt{10}$ MeV. Two different truncations on the $j$ index in the Eq.~(2) are used ($i\leq 2$ in both cases): $j\leq 3$ and $j\leq 2$, corresponding to ``Up to $E^3$'' and ``Up to $E^2$'' results. The shaded region indicates the range of the Eq.~(2)'s RHS with $E$ above 5 MeV.}\label{fig:checkLepagePlot}
\end{figure*}
However, the truncation on the series sum in the improved BERW formula inevitably leads to truncation error. Fig.~\ref{fig:checkLepagePlot} plots such errors, i.e., the difference between the formula's two sides with the $j$ index summed up to either $3$ (for the ``Up to $E^3$'' results) or $2$ (for the ``Up to $E^2$'' results) and with $i$ index summed up to $2$ for both cases. The shaded region shows the rough range of the formula's RHS with $E$ above 5 MeV. In the low energy region, the error does behave like the leading terms left out of the series sum, but it increases with $E$ and eventually reaches the shaded region. The latter signals the break down of the series expansion, as the error reaches $100\%$. This also suggests the break down scale for $E$ is above $20$ MeV and perhaps below 40 MeV in all three channels, which are consistent with my choice of $\MH=200$ MeV. Note the behavior of the error above the break down scale could be more complicated than the simple power law in the low-energy region. In fact in all these channels, the errors oscillate and change their signs in the higher-energy range not shown in the plots.

\section{Details on the renormalization in the EFT derivation}
This section discusses the particular renormalization scheme mentioned in the main text, in which the divergences in in $\Sigma$ and $\Sigma_\omega$ cancel in Eq.~(9) in the main text. This scheme gives a specific relation between $\Lambda$ and $n_\Lambda$.  To derive this relation, the $n_\Lambda$-dependence of $\Sigma_\omega$ needs to be studied. 

 Two formulas are useful for understanding $f_\ell(z_E, n)$ at large $n$ and $\Sigma_\omega$ at large $n_\Lambda$. 
 The first is~\cite[Eq.~5.11.13]{NIST:DLMF} 
\begin{eqnarray}
\frac{\Gamma(z+a)}{\Gamma(z+b)} & \stackrel{z\rightarrow\infty}{\sim} &  \sum_{k=0}^{+\infty} \frac{G_k(a,b)}{z^{k-a+b}}  \ \text{if}\, \arg(z)\leq \pi - 0^+ .  \label{eqn:gammaratioasym} 
\end{eqnarray} 
Here $a$ and $b$ are real or complex constants, and $G_k(a,b)$ as a function of $a$ and $b$ is related to the generalized Bernoulli polynomials~\cite[Eq.~5.11.17]{NIST:DLMF}~\cite{Temme1996}. 
The second is the Euler–-Maclaurin formula~\cite[Eq.~2.10.1]{NIST:DLMF}, stating that for a smooth $f(x)$, its series sum can be approximated using an asymptotic expansion:
\begin{eqnarray}
\!\!\!\!\!\!\!\!\sum_{n}^{n_\Lambda}\! f(n) \sim\!\! \int\displaylimits^{n_\Lambda} \! f(x) dx  \!+\! \frac{f(n_\Lambda)}{2} +\sum_{j=1}^{+\infty} \frac{B_{2j} }{(2j)!}\!\frac{d^{{2j-1}}\! f(n_\Lambda) }{d n_\Lambda^{2j-1}}, \label{eqn:EM}
\end{eqnarray}
where $B_n$ is a Bernoulli number.
Only the $n_\Lambda$ dependent terms are shown.
 Like $\Sigma(E)$, the $0^{\textrm{th}}$ to $\lth$ derivatives of $\Sigma_\omega(E)$ diverge. 
 So for s-waves,  only  $\Sigma_\omega(0)$ is considered: 
\begin{eqnarray}
\!\!\!\!\!\!\!\!\! \sum_{n}^{n_\Lambda}\!f_{\ell=0}\left(0,\! n\right)\! 
=\!\!\sum_{n}^{n_\Lambda} \!\frac{-1}{\sqrt{n}}\!\!\left[1 \!+\! O(\frac{1}{n})\right] \!\sim\!  - 2 n_\Lambda^\half\!+\! O( n_\Lambda^{-\half}). 
\end{eqnarray}
Thus $\bar{T}_\Lambda = 2 n_\Lambda (1+ O(n_\Lambda^{-1}))$ ($\bar{T}_\Lambda\equiv T_\Lambda/\omega$) so that the divergence can be absorbed by $\sqrt{2 \bar{T}_\Lambda}$ in the main text's Eq.~(10). 
The $O(n_\Lambda^{-1})$ term in the $n_\Lambda$-$\Lambda$ relation is not relevant when $n_\Lambda \to \infty$. 
For p-waves, the $0^{\textrm th}$ and first derivatives are divergent:  
\begin{eqnarray}
&& \sum_{n}^{n_\Lambda}f_{\ell=1}\left(0,n\right)  \sim   -\frac{2}{3} n_\Lambda^{\frac{3}{2}} - \frac{7}{4} n_\Lambda^{\half}  + O\left(n_\Lambda^{-\half}\right) ,  \label{eqn:divdwave1} \\ 
&& \sum_{n}^{n_\Lambda} \partial_{\ze}f_{\ell=1}\left(\ze,n\right)\vert_{\ze=0}
 \sim -2 n_\Lambda^{\half}   + O\left(n_\Lambda^{-\half}\right).   
\end{eqnarray}
Thus, ${\bar{T}_{\Lambda}} = 2 n_\Lambda \left(1+\frac{7}{4}  n_\Lambda^{-1} + O(n_\Lambda^{-2})\right) $, to have these divergences canceled by those in $\mathcal{P}\Sigma(E)$ in the main text's Eq.~(10). 
The $\frac{7}{4}  n_\Lambda^{-1}$ piece in the expression must be specified, but higher-order terms are not needed. 
However for d-waves, another order higher needs to be specified: $\bar{T}_{\Lambda} = 2 n_\Lambda \left(1+\frac{9}{4}  n_\Lambda^{-1}- \frac{37}{32}  n_\Lambda^{-2} + O(n_\Lambda^{-3})\right)$; for even larger $\ell$, more terms need to be specified accordingly.  

This requirement is tied to the fact that for a specific $E$-derivative, there is a tower of divergences with different degrees in $\Sigma_\omega$ (e.g.,~\eq{eqn:divdwave1}), in stark contrast with the same derivative of $\Sigma$, where the power of $\Lambda$ is fixed by the the dimensions.

\section{A proof of Eq.~(11) in the main text} 
\begin{figure}[h!]
\includegraphics[width=0.3\textwidth]{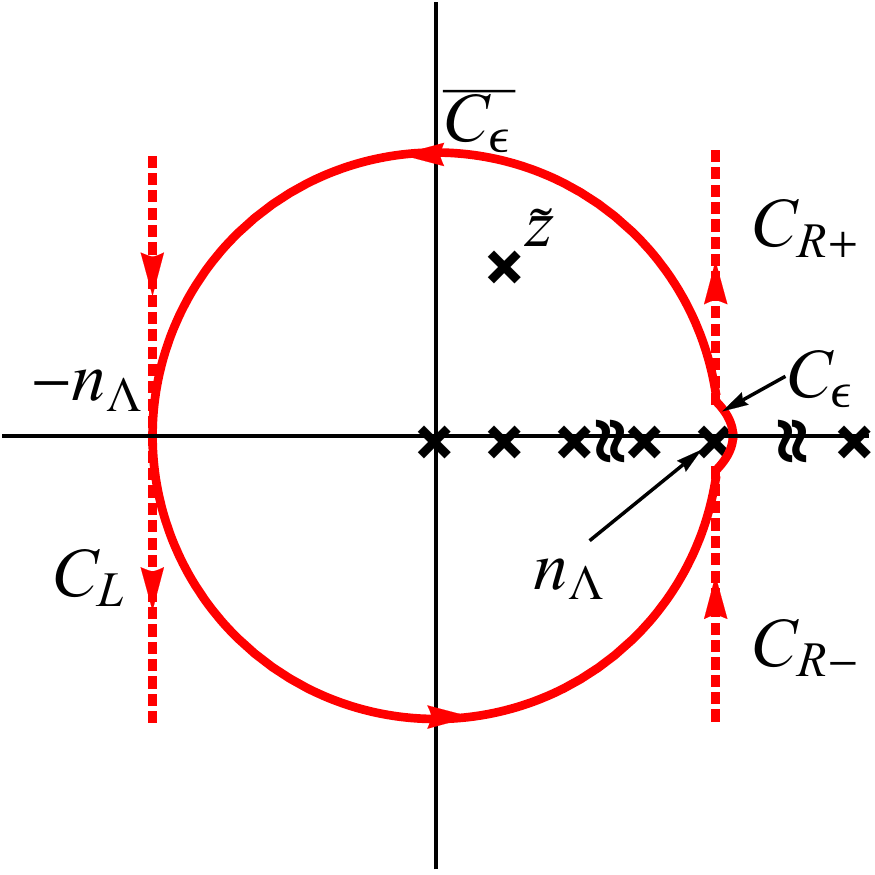}
\caption{The contours for the integration in the complex $\tilde{u}$ plane. $C_\epsilon$ is a small semi-circle around the pole at $\tilde{u}=n_\Lambda$ with radius $\epsilon$, which intersects with a large circle $\overline{C_\epsilon}$ around the center with radius $R=n_\Lambda$. The $C_{R\pm}$ originates from the intersections and goes from  $n_\Lambda-i\infty$ and to $n_\Lambda+i\infty$ in parallel to the imaginary axis, while $C_L$ is also parallel with the imaginary axis and runs from $-n_\Lambda+i\infty$ to $-n_\Lambda-i\infty$.}\label{fig:intcontour2}
\end{figure}
Let us redefine $\tilde{z} \equiv z -\ell/2-3/4$, and 
\begin{eqnarray}
 \tilde{f}_\ell(\tilde{z},n) & \equiv &  f_\ell(\tilde{z} +\frac{\ell}{2}+\frac{3}{4},n)= \frac{ {\Gaf{n + \ell+ \threehalf}}}{\Gaf{n + 1}(\tilde{z}-n)}   \ , \label{eqn:defftilde}\\
 g_\ell(\tilde{z})& \equiv& (-)^\ell\pi \frac{\Gaf{-\tilde{z}}}{\Gaf{-\ell-1/2-\tilde{z}}} \ . \label{eqn:redefgl}
\end{eqnarray}
The main text's Eq.~(11) then becomes  
\begin{eqnarray}
\sum_{n=0}^{(\mathcal{R})} \tilde{f}_\ell(\tilde{z},n) = g_\ell(\tilde{z}) \ . \label{eqn:conj3}
\end{eqnarray}
The renormalization $\mathcal{R}$ is defined by Eq.~(10) in the main text
and the relationship between $T_\Lambda$ and $n_\Lambda$ that guarantees the cancellation of the divergences on that equation's left side. 
The proof starts with integrating $g_\ell(\tilde{u})/(\tilde{u}- \tilde{z})$ in the  complex $\tilde{u}$ 
plane over a large contour around the origin and crossing between the two singularities 
at $\tilde{u}=n_{\Lambda}$ and $ n_{\Lambda}+1$ on the positive real axis. The contour $\overline{C_\epsilon}+C_\epsilon$ is plotted in Fig.~\ref{fig:intcontour2}. 
After rearranging terms one get
\begin{eqnarray} 
g_\ell(\tilde{z})&=& \sum_{n=0}^{n_\Lambda} \tilde{f}_\ell(\tilde{z}, n) + \frac{1}{2\pi i} \oint\displaylimits_{\overline{C_\epsilon}+C_\epsilon} d\tilde{u} \frac{g_\ell(\tilde{u})}{\tilde{u}- \tilde{z}}\ .  \label{eqn:contour2}
\end{eqnarray}
The left side comes from the residue of the $1/(\tilde{u}-\tilde{z})$ pole in the contour integration
 while the series sum on the right side is from the residues of $g_\ell(\tilde{u})$'s poles 
 (all on the positive real axis) in the same integration.
 Comparing \eq{eqn:contour2} to \eq{eqn:conj3} suggests that the contour integration in \eq{eqn:contour2} should cancel the series's divergence as the $\bar{T}_\Lambda$ terms---i.e., those from $\mathcal{P}\Sigma(E)$---do in the main text's Eq.~(10).
 This is the focus of the following proof. 

As preparation, it is important to understand the behavior of $g_\ell(\tilde{u})$ in two 
different regions (define $\theta_{\tilde{u}}\equiv \arg(\tilde{u})$ and $R$ as the 
radius of $\overline{C_\epsilon}$): $ \pi \geq |\theta_{\tilde{u}}|\gg R^{-1} $ 
and $|\theta_{\tilde{u}}| \lesssim R^{-1} $. 
Let us look at the region close to the positive real axis first. 
Considering the presence of $g_\ell(\tilde{u})$'s poles,  it is easier to use the following re-expression based on \cite[Eq.~5.5.3]{NIST:DLMF}: 
\begin{eqnarray}
g_\ell(\tilde{u}) = \pi \cot(\pi\tilde{u}) \frac{\Gaf{\ell+\frac{3}{2}+ \tilde{u} }}{\Gaf{1+ \tilde{u} }} \ . \label{eqn:glredef}
\end{eqnarray}
This expression moves the poles from the $\Gamma$ function to the $\cot$ function.   When $R\rightarrow +\infty$, 
\begin{eqnarray}
\cot(\pi\tilde{u})&=& i \frac{e^{i\pi\tilde{u}}+ e^{-i\pi\tilde{u}} }{e^{i\pi\tilde{u}}- e^{-i\pi\tilde{u}} } \!=\!
\begin{cases}
\!(- i),  \pi-\theta_0 >\theta_{\tilde{u}}\gg \frac{1}{R} \\ 
\!(+ i),   \theta_0-\pi < \theta_{\tilde{u}}\ll \!\!-\frac{1}{R} \\ 
\end{cases} \label{eqn:cotasym}
\end{eqnarray}
Here $\theta_0$ is a small positive number. Therefore, $\cot(\pi\tilde{u}) \sim -i \sgn(\text{Im\,}\tilde{u})$ when $ \pi-\theta_0 > |\theta_{\tilde{u}}|\gg R^{-1}$ [the error scales as $\exp(-2\pi |\text{Im\,}(\tilde{u})|)$], but in the \second\ region, $|\theta_{\tilde{u}}| \lesssim R^{-1}$, and its behavior is qualitatively different. 

Since the $\Gamma$ function ratio in \eq{eqn:glredef} is the same as that ratio in $\tilde{f}_\ell(\tilde{z}, n)$ with $n \rightarrow \tilde{u}$, by applying the asymptotic expansion from the main text's Eq.~(11) 
on the right side of \eq{eqn:glredef}, I get, when $R\rightarrow +\infty $ such that $R \gg |\tilde{z}|$ and when $ \pi-\theta_0 > |\theta_{\tilde{u}}|\gg R^{-1}$, 
\begin{eqnarray}
\frac{g_\ell(\tilde{u})}{\tilde{u}-\tilde{z}} &\sim & - i\pi \sgn(\text{Im\,}\tilde{u}) \left(\sum_{k=0}^{+\infty}  \frac{G_k(\ell+\frac{3}{2}, 1)}{{\tilde{u}}^{k-\ell-\half}} \right) \sum_{k'=0}^{+\infty}\frac{\tilde{z}^{k'}}{\tilde{u}^{k'+1}} \notag \\ 
&=& i \pi \sgn(\text{Im\,}\tilde{u})  \tilde{f}_\ell^\mathrm{E}(\tilde{z},\tilde{u}). \label{eqn:glasymp}
\end{eqnarray}
Here $\tilde{f}_\ell^\mathrm{E}(\tilde{z}, \tilde{u}) $ is the asymptotic expansion of $\tilde{f}_\ell(\tilde{z},\tilde{u}) $ in terms of $1/\tilde{u}$ using the main text's Eq.~(11). 
It turns out the above expansion also holds when $\theta_u \rightarrow \pi$ or $-\pi$. However, using \eq{eqn:glredef} to understand $g_{\tilde{u}}$ close to the negative real axis is awkward, as it involves the cancellation of poles from the $\cot$ and the $\Gamma$ functions. Instead, 
Eq.~(11) in the main text can be applied to analyze the original form of $g_\ell(\tilde{u})$ (see \eq{eqn:redefgl}): 
\begin{eqnarray}
g_\ell(\tilde{u}) &\sim& (-)^\ell \pi (-\tilde{u})^{\ell+\half}  \sum_{k=0}^{+\infty}  \frac{G_k\left(0,-\ell -\half \right)}{(-\tilde{u})^k} \notag \\ 
&=& - i \sgn(\text{Im\,}\tilde{u})  \pi\,  \tilde{u}^{\ell+\half}  \sum_{k=0}^{+\infty}  \frac{G_k(\ell+\frac{3}{2}, 1)}{{\tilde{u}}^k}  \;. 
\end{eqnarray}
Here $G_k(\ell+\frac{3}{2}, 1) = (-)^k G_k(0, -\ell-\half)$ is used, which can be inferred from its definition~\cite[Eq.~5.11.17]{NIST:DLMF}~\cite{Temme1996} and properties of the generalized Bernoulli polynomials. 
It is worth pointing out that the $\sgn(\text{Im\,}\tilde{u})$ factor moves the asymptotic series's branch cut on the negative real axis due to the $\sqrt{\tilde{u}}$\footnote{As usual~\cite[Eq.~1.9.7]{NIST:DLMF}, the $\sqrt{\tilde{u}}$ branch cut in the complex $\tilde{u}$ plane is on the negative real axis. Meanwhile $\Sigma(E)$ (and the related scattering $T$-matrix), as calculated in the main text's Eq.~(5), 
in the complex $E$ plane has a branch cut separating physical and unphysical sheets on the positive real axis~\cite{GoldbergerQM}, because it involves $-i p = \sqrt{-(2 \mr E + i0^+)}$.}  factor to the positive real axis, ensuring that the expansion series is analytic around the negative real axis. Therefore, \eq{eqn:glasymp} holds  when  $ \pi \geq |\theta_{\tilde{u}}|\gg R^{-1}$, but not in the $|\theta_{\tilde{u}}| \lesssim R^{-1} $ region. This suggests splitting the contour integration into two major pieces: 
\begin{eqnarray}
&& \overbrace{ \oint\displaylimits_{\overline{C_\epsilon}+C_\epsilon-S}\!\!\!   i \pi \sgn(\text{Im\,}\tilde{u})  \tilde{f}_\ell^\text{E}(\tilde{z}, \tilde{u})  \frac{d\tilde{u}}{2\pi i} }^{\mathrm{1^{st}\ piece}} \notag \\ 
+ &&  \overbrace{ \oint\displaylimits_{\overline{C_\epsilon}+C_\epsilon-S}\!\!\! \frac{d\tilde{u}}{2\pi i} \left[\frac{g_\ell(\tilde{u})}{\tilde{u}-\tilde{z}}   -   i \pi \sgn(\text{Im\,}\tilde{u})  \tilde{f}_\ell^\text{E}(\tilde{z}, \tilde{u}) \right]   }^{\mathrm{2^{nd}\ piece}} .  \label{eqn:largeContour1} 
\end{eqnarray}
In $\overline{C_\epsilon}+C_\epsilon-S$, $S$ is the point where $C_\epsilon$ crosses the real axis. 
Thus the contour means to integrate infinitely close to the upper and lower real axis, because of the corresponding integrand's  branch cut on the positive real axis. 
In the \second\ piece, the first term should be integrated over $\overline{C_\epsilon}+C_\epsilon$. However, since the integrand is continuous across the point $S$ as long as $\epsilon \neq 0$, changing the contour to $\overline{C_\epsilon}+C_\epsilon-S$ does not affect the results. 

Integrating the \first\ piece over $\overline{C_{\epsilon}}$ with $\epsilon\rightarrow 0^+$ gives 
\begin{eqnarray}
&& \int\displaylimits_{ 0<\theta_{\tilde{u}}  <2\pi} \!\!\!\! d\tilde{u}\, \frac{ \tilde{f}_\ell^\text{ET}(\tilde{z}, \tilde{u})}{2} 
= \int\displaylimits_{ 0< \theta_v<\pi} dv\,  v  \tilde{f}_\ell^\text{ET}(\tilde{z}, v^2)  \notag \\   
&& =\int\displaylimits_{-\sqrt{R}}^{\sqrt{R}} dv  (-)  v  \tilde{f}_\ell^\text{ET}(\tilde{z}, v^2) = (-) \int\displaylimits_{0}^{R} d\tilde{u}   \tilde{f}_\ell^\text{ET}(\tilde{z}, \tilde{u})
  \;. 
\end{eqnarray}
Here $ \tilde{f}_\ell^\text{ET}(\tilde{z}, \tilde{u})$ truncates the expansion in $ \tilde{f}_\ell^\text{E}(\tilde{z}, \tilde{u})$ by only keeping 
 terms with powers of $\tilde{u}$ from $\ell-1/2$ to $-1/2$ (the neglected terms' integration vanishes no slower than $O(1/\sqrt{R})$ as $R\rightarrow \infty$). In the derivation, (1) the $-\pi\leq \theta_{\tilde{u}}< 0$ branch has been rotated by $+2\pi$, which eliminated the $\sgn(\text{Im\,}\tilde{u})$ factor because of the $\sqrt{\tilde{u}}$ factor in the integrand; (2) a transformation, $\tilde{u}=v^2$, was used; and (3) the fact that $v  \tilde{f}_\ell^\text{E,T}(\tilde{z} v^2)$ is analytic in the upper complex $v$ plane and an even function on the real axis was used. Note that integrating the \first\ piece over $C_{\epsilon}-S$ gives $0$ with $\epsilon\rightarrow 0^+$. 

For the \second\ piece, the $\overline{C_\epsilon}+C_\epsilon - S$ contour can be deformed to $C_L+C_{R-}+C_\epsilon+C_{R+}-S$ without crossing any singularities,  with the left and right segments connecting at $\infty$. The integrand on the two sets of contours (including in the area enclosed by them) is 0 up to at most a $\sim e^{-R \#}$ correction (with $\#$ as a positive number), except on the segments with $|\theta_{\tilde{u}}| \lesssim 1/R $ along $C_{R\pm}$ and $C_\epsilon$. Therefore, $C_L$ can be safely ignored. Let us focus on $C_{R-}+C_\epsilon+C_{R+}-S$. Since  
\begin{align}
\frac{g_\ell(\tilde{u})}{\tilde{u}-\tilde{z}} \sim - \pi \cot(\pi\tilde{u}) \tilde{f}_\ell^\text{E}(\tilde{z}, \tilde{u})  \ , \notag  
\end{align}
 the \second\ piece is 
\begin{align}
\int\displaylimits_{C_{R\pm}+C_\epsilon-S} \!\!\!  \frac{d\tilde{u}}{2\pi i}  
(-\pi)\left[ \cot(\pi\tilde{u})+ i \sgn(\text{Im\,}\tilde{u}) \right]  \tilde{f}_\ell^\text{E}(\tilde{z}, \tilde{u}). 
\end{align}
Note that $\cot(\pi\tilde{u}) \sim 1/[\pi (\tilde{u}-n_\Lambda)]$ when $\tilde{u} \rightarrow n_\Lambda$, its integration over $C_\epsilon - S$  with $\epsilon\rightarrow 0^+$ gives 
$(-)\tilde{f}_\ell^\text{E}(\tilde{z}, n_\Lambda )/2 $. 

Integrating over $C_{R\pm}$ with $\epsilon\rightarrow 0^+$ gives   
\begin{align}
& \int\displaylimits_{C_{R+} + C_{R-}} d\tilde{u} \left[\cot(\pi\tilde{u})+ i \sgn(\text{Im\,}\tilde{u})\right] (-) \frac{\tilde{f}_\ell^\mathrm{E}(\tilde{z}, \tilde{u})}{2i}   \notag \\ 
=  & \int_{\epsilon}^{+\infty} i\,d\Delta \frac{ \tilde{f}_\ell^\mathrm{E}(\tilde{z}, n_{\Lambda}+i\Delta) }{e^{2\pi\Delta}-1} - \int^{-\epsilon}_{-\infty} i\,d\Delta \frac{  \tilde{f}_\ell^\mathrm{E}(\tilde{z}, n_{\Lambda}+ i\Delta) }{e^{-2\pi\Delta}-1}  \notag \\ 
= & \int_{0}^{+\infty} \frac{d2\pi\Delta}{2\pi i}  \frac{ \left[\tilde{f}_\ell^\text{E}(\tilde{z}, n_\Lambda- i\Delta) - \tilde{f}_\ell^\text{E}(\tilde{z}, n_\Lambda+ i\Delta)\right]}{e^{2\pi\Delta}-1} \notag \\ 
= &(-) \sum_{j=1}^{+\infty}\frac{\partial^{2j-1}\! \tilde{f}_\ell^{\text{E}}\!(\tilde{z},\! n_\Lambda)}{\partial n_\Lambda^{2j-1}} \frac{B_{2j}}{(2j)!} \ .
\end{align}

Because the integration is dominated by the $\Delta \ll n_\Lambda$ region, the two $\tilde{f}_\ell^\mathrm{E}$ are expanded in terms of Taylor series for the \second\ argument at $n_\Lambda$ in the above derivation. The resulted integrations are proportional to Riemann $\zeta(2j)$ at even arguments, which are related to Bernoulli numbers \cite[Eq.~25.5.1, 25.6.2, 24.2.2]{NIST:DLMF}. Adding all the contributions, one can see that the \second\ term on the right side of \eq{eqn:contour2} can be approximated by  
\begin{align} 
\!\! -\!\!\int\displaylimits_{0}^{n_\Lambda} 
   \! \tilde{f}_\ell^{\text{ET}}\!(\tilde{z}, \tilde{u})d\tilde{u} 
   - \frac{\tilde{f}_\ell^{\text{E}}(\tilde{z}, n_\Lambda)}{2}  
   - \!\sum_{j=1}^{+\infty}\! \frac{B_{2j}}{(2j)!} 
   \frac{\partial^{2j-1}\!\tilde{f}_\ell^{\text{E}}(\tilde{z},\! n_\Lambda)}{\partial n_\Lambda^{2j-1}} 
     \;, \notag 
\end{align} 
with the leading error scaling as $n_\Lambda^{-\half}$ [see the discussion of $\tilde{f}_\ell^\text{ET}$. In comparison, the $e^{-n_\Lambda\#}$ error ($\#$ is a positive constant) is much smaller with large $n_\Lambda$]. The above expression is exactly the same as the divergent $n_\Lambda$-dependent pieces in $\Sigma_\omega$'s series sum derived using \eq{eqn:EM} in the previous section, 
which completes the proof.

\end{document}